# Autoreactivity to malondialdehyde-modifications in rheumatoid arthritis is linked to disease activity and synovial pathogenesis

Caroline Grönwall[a,b]*, Khaled Amara[a], Uta Hardt[a], Akilan Krishnamurthy[a], Johanna Steen[a], Marianne Engström[a], Meng Sun[a], A. Jimmy Ytterberg[a,c], Roman A. Zubarev[c], Dagmar Scheel-Toellner[d], Jeffrey D. Greenberg[b], Lars Klareskog[a], Anca I. Catrina [a], Vivianne Malmström[a] and Gregg J. Silverman[b]

a) Rheumatology Unit, Department of Medicine, Karolinska Institutet and Karolinska University Hospital, Stockholm, Sweden
b) Division of Rheumatology, Department of Medicine, NYU School of Medicine, New York, NY, USA
c) Department of Medical Biochemistry and Biophysics, Karolinska Institutet, Stockholm, Sweden
d) Rheumatology Research Group, Centre for Translational Inflammation Research, College of Medical and Dental Sciences, University of Birmingham, Birmingham, United Kingdom

**Corresponding author:** Caroline Grönwall, *Department of Medicine, Rheumatology Unit, Karolinska Institutet and Karolinska University Hospital, Center for Molecular Medicine L8:04, 17176 Stockholm, Sweden.*
E-mail: caroline.gronwall@ki.se

## Abstract

Oxidation-associated malondialdehyde (MDA) modification of proteins can generate immunogenic neo-epitopes that are recognized by autoantibodies. In health, IgM antibodies to MDA-adducts are part of the natural antibody pool, while elevated levels of IgG anti-MDA antibodies are associated with inflammatory and autoimmune conditions. Yet, in human autoimmune disease IgG anti-MDA responses have not been well characterized and their potential contribution to disease pathogenesis is not known. Here, we investigate MDA-modifications and anti-MDA-modified protein autoreactivity in rheumatoid arthritis (RA). While RA is primarily associated with autoreactivity to citrullinated antigens, we also observed increases in serum IgG anti-MDA in RA patients compared to controls. IgG anti-MDA levels significantly correlated with disease activity by DAS28-ESR and serum TNF-alpha, IL-6, and CRP. Mass spectrometry analysis of RA synovial tissue identified MDA-modified proteins and revealed shared peptides between MDA-modified and citrullinated actin and vimentin. Furthermore, anti-MDA autoreactivity among synovial B cells was discovered when investigating recombinant monoclonal antibodies (mAbs) cloned from single B cells, and 3.5 % of memory B cells and 2.3% of plasma cells were found to be anti-MDA positive. Several clones were highly specific for MDA-modification with no cross-reactivity to other antigen modifications such as citrullination, carbamylation or 4-HNE-carbonylation. The mAbs recognized MDA-adducts in a variety of proteins including albumin, histone 2B, fibrinogen and vimentin. Interestingly, the most reactive clone, originated from an IgG1-bearing memory B cell, was encoded by near germline variable genes, and showed similarity to previously reported natural IgM. Other anti-MDA clones display somatic hypermutations and lower reactivity. Importantly, these anti-MDA antibodies had significant *in vitro* functional properties and induced enhanced osteoclastogenesis, while the natural antibody related high-reactivity clone did not. We postulate that these may represent distinctly different facets of anti-MDA autoreactive responses.

**Highlights:**

* Malondialdehyde modification of self-proteins occurs in the RA joint
* IgG anti-MDA reactivity is elevated in RA and associated with disease activity
* Specific anti-MDA reactive synovial B cells have been recovered from RA patients
* Different classes of IgG anti-MDA reactivity exists
* Some anti-MDA IgG enhance osteoclastogenesis

## 1. Introduction

Rheumatoid arthritis (RA) is a chronic and potentially disabling autoimmune disease affecting between 0.5 and 1% of the Western population [1]. RA is associated with synovial inflammation and progressive destruction of joints. There is also increased risk for morbidity and mortality from accelerated atherosclerotic cardiovascular disease [2]. Pathogenesis is associated with a characteristic autoimmune response to self-proteins post-translationally modified by citrullination, resulting in circulating anti-citrullinated protein antibodies (ACPA) that are detected in 65-80% of patients with established disease [3] (reviewed in [4]).

Seropositive RA is defined by clinical criteria that include the presence of IgG antibodies to synthetic peptides, termed cyclic citrullinated peptides (CCP), and/or IgM rheumatoid factors (RF) that bind aggregated IgG Fc regions [5]. Yet RA patients also commonly display other autoantibodies, including IgG binding to proteins post-translationally modified by carbamylation, that involves the generation of homocitrulline residues from lysines [6, 7]. Antibodies to carbamylated antigens have been reported in both seropositive and seronegative RA patients. They can arise in parallel to ACPA responses and could at times reflect unique epitope recognition, but may also be partly explained by ACPA cross-reactivity [8].

The current report focuses on the contribution of immune responses to a distinctly different and less extensively studied self-protein modification, malondialdehyde (MDA) adducts, during RA pathogenesis.
MDA is a naturally occurring, highly reactive aldehyde, produced under oxidative stress states associated with excessive generation of reactive oxygen species (ROS). Elevated ROS catalyzes membrane lipid peroxidation and the formation of reactive MDA that can covalently modify proteins through carbonylation of amino acids carrying free amine groups (i.e. lysine, arginine, histidine), and to less extent amino acids with amide groups (i.e. asparagine, glutamine), which can generate structural changes and neo-epitopes [9, 10]. A large number of self-proteins have been found to be modified by MDA under local inflammatory conditions, including vimentin, fibrinogen, $\alpha$-enolase and albumin [11, 12] (reviewed in [13, 14]). Acetaldehyde (AcA) can further react with MDA adducts to form immunogenic malondiadehyde-acetaldehyde (MAA) modifications (Figure 1) [15]. ROS levels sufficient to cause tissue injury can be generated by exogenous stimuli such as tobacco smoke, or be endogenously produced during inflammation [16]. Moreover, oxidized proteins and lipids are also formed as a consequence of programmed death pathways and the resulting adducts on apoptotic cells can be recognized by some anti-MDA antibodies [17-19].
Recent evidence implicates that autoimmune diseases are associated with an altered redox-state and elevated levels of oxidative species, which may contribute to disease pathogenesis [20-25]. Increased levels of MDA and MDA-modified proteins, observed both in systemic lupus erythematosus (SLE) and RA, may reflect disturbances in oxidation balance occurring during systemic inflammation [20-25]. Strikingly, at birth the human natural IgM repertoire has a strong bias towards antibody-recognition of oxidation-associated epitopes, especially MDA-modifications [18, 19, 26]. Hence, the immune system is primed from birth to recognize modified self-antigens, and it has been postulated that these antibodies play important roles in clearance of apoptotic cells, neutralization of harmful molecules, and maintenance of immune homeostasis [27]. However, whilst IgM may be protective, the constant regions of IgG autoantibodies with the same specificity may instead trigger inflammatory responses. IgG autoantibodies that bind MDA can be elevated in SLE, and levels are directly related to increased disease activity [25, 28].
Herein, we have investigated the representation of MDA-modifications of self-proteins in the rheumatoid joint and their potential as immunogens for eliciting immune responses. In a cross-sectional RA cohort, we also evaluated serological levels of autoantibodies to MDA-modified self-protein epitopes in relation to disease activity. Furthermore, we recovered human mAbs isolated from individual synovial B cells and plasma cells from RA patients, and evaluated their binding specificity for MDA-modified self-proteins and their associated *in vitro* functional properties.

## 2. Material and Methods
### 2.1 Patients and sample procedures
All RA patients fulfilled the 2010 ACR/EULAR criteria for diagnosis [5], and informed consent was obtained for all patients and controls according to protocols approved by the Human Subjects Institutional Review Board of NYU School of Medicine, the Ethics Review Committee at the University of Birmingham, and the Ethics Review Committee North at the Karolinska University Hospital. Patients were classified as seropositive RA and seronegative RA based on the clinical CCP2 assay. Synovial fluid samples for cell isolation were collected in connection to when the patients required arthrocentesis due to local disease activity, and the patients were given subsequent local steroid injections. Synovial tissues of RA patients or disease controls were obtained at the time of joint-replacement surgery.

### 2.2 Mass spectrometry analysis
*In vitro* MDA-modified BSA was analyzed in pilot studies for MDA-detection and for comparison between different MDA-modification times. Ten µg protein was reduced and alkylated by DTT and iodoacetamide (Sigma Aldrich), precipitated and digested by trypsin (sequencing grade,

Promega) in 50 mM ammonium bicarbonate and 30% DMSO (Sigma Aldrich) at protease to protein concentration of 1:20, as previously described [29]. After desalting the peptides using C18 StageTips (Thermo Fisher Scientific), the peptides were analyzed by nanoLC-MS/MS a NanoUltimate 3000 coupled on-line to an LTQ Orbitrap Velos mass spectrometer, were the Velos Orbitrap had been upgraded to an Elite (Thermo Fisher Scientific, Germany). The peptides were separated using a Acclaim® PepMap100 precolumn (C18, 3 μm, 100 Å; Thermo Scientific) together with a 15 cm EASY-Spray PepMap® analytical column (C18, 3μm, 100Å; Thermo Scientific). The separation was achieved using ACN/water gradients (buffer A: 2% ACN, 0.1% FA; buffer B: 98% ACN, 0.1% FA) of 5–26% B over 55 min, followed by a 26–95% ACN gradient over 5 min and 95% ACN for 8 min, all at a flow rate of 300 nl/min. The instruments were operated in a data-dependent mode with a top 5 method. The mass spectra were acquired at a resolution of 60,000 followed by either CID only or HCD only MS/MS fragmentation. A normalized collision energy of 35 was used for CID and 30 for HCD. The HCD MS/MS spectra were acquired at a resolution of 15,000. One pmol of the three samples were analyzed by both MS analysis using CID only and HCD only, each in two technical replicates (i.e. four analyses per samples).

Data from an earlier reported study of post-translational citrullination in rheumatoid synovial tissue [30], were re-searched for the detection of MDA-modified proteins. As primary analyses sought to detect citrullinated peptides the samples were enzymatically processed with Lys-C to avoid digestion at arginine sites. Note that this will reduce the detection of lysine modified residues. As previously described, each synovial tissue sample was analyzed by four different MS methods: top5 using CID/ETD fragmentation, top5 using ETD only, top4 using HCD only, and top4 using an inclusion list together HCD only (the inclusion list contained citrullinated peptides identified in the same study). The 28 raw files [30] were re-processed by Raw2MGF v2.1.3 and re-searched against the human complete proteome database (downloaded from www.uniprot.org April, 2013; 71434 sequences; 24507501 residues). At least one spectrum from each peptide reported in the supplementary Table 1 and 2 was validated manually.

## 2.3 Single cell cloning of synovial B cells and preparation of human monoclonal antibodies

Monoclonal antibodies (mAbs) from single synovial memory B cells or plasma cells were generated as recently described [31-33]. Notably, the current study includes additional patient samples than previously published. Briefly, cryopreserved synovial mononuclear cells were thawed, stained with specific fluorescently labeled antibodies, and flow cytometric sorted as either; single CD19 FcRL4+/- B cells [31, 34], or CD19+ IgG+ cells [32], into 96-well plates containing 5 μl per well of 0.5x PBS 10 mM DTT and RNAsin (RNAase inhibitor; Promega). For isolation of single synovial antibody secreting cells (plasmablasts/plasma cells), the fluorescent foci method [35] was applied using anti-human IgG specific beads and FITC labeled anti-human IgG (gamma-specific), to identify all IgG secreting cells in the synovial sample by fluorescent microscopy [33]. Cells that displayed an IgG fluorescent halo were extracted using an Eppendorf NK micromanipulator, and single-cell cloning was then performed, with the same protocol used for flow cytometry sorted memory B cell [32]. Immunoglobulin variable genes were cloned into human heavy- and light-chain IgG expression vectors, after cDNA synthesis and PCR amplification, using established methods [32, 36]. Recombinant mAbs were expressed as IgG1 in the Expi293 system (Thermo Fisher Scientific) with transient transfection using PEI-max (total 38 μg plasmid DNA to 30 ml cells, IgH and IgL constructs). Antibodies were purified using protein G affinity purification (Protein G Fast Flow Sepharose, GE Healthcare Life Sciences) and their concentrations determined by IgG ELISA. Selected mAbs were also produced in larger scale (i.e. 400-3200 ml cultures), with purified products subsequently characterized with quality testing that included SEC analysis for aggregation, SDS-PAGE, specificity ELISA, and endotoxin testing.

## 2.4 Antigen-specificity assays for serum and monoclonal antibodies

Serological screenings of specific IgG and IgM antibody levels in clinical samples were performed using previously reported methods [28, 37]. Wells were coated with MDA-BSA (Academy Bio-Medical) or PC-BSA (low load, Biosearch Technologies) at 3 μg/ml, blocked with 3% BSA in PBS, and serum samples were diluted 1:200 and 1:1000. IgG, IgA or IgM reactivity was detected with HRP-conjugated gamma-specific goat (Fab')2 anti-human IgG, goat anti-human IgA (Jackson ImmunoResearch) or mu-specific goat anti-human IgM (Southern Biotech), and developed with TMB substrate (Biolegend). All absorbance values were quantified towards a serum reference sample and presented as Relative Units (RU)/ml. Serum rheumatoid factor and anti-CCP3 levels were determined using clinical assays, (QUANTA lite RF IgM and CCP3), according to the manufacturer's instructions (Inova Diagnostics). For IgM and IgA anti-CCP3 reactivity, plates were coated with CCP3 peptide at 3 μg/ml (gift from Inova Diagnostics) and reactivity was detected using HRP-conjugated goat anti-human IgM (Southern Biotech) or goat anti-human IgA (Jackson ImmunoResearch).

Similarly, mAbs were screened for MDA reactivity at 3 ug/ml using an adapted assay from above. All mAbs were also screened for non-specific binding to unmodified BSA (molecular grade BSA, NEB) and RF activity using wells coated with chromopure rabbit IgG (Jackson ImmunoResearch), and binding detected with HRP-conjugated rabbit (Fab')2 gamma-specific anti-human IgG (Jackson ImmunoResearch). Reactivity of selected clones to other antigens, carbamylated BSA, 4-HNE-modified BSA, MDA-LDL, LDL (Academy Bio-Medical) and PC-BSA (Biosearch Technologies), were similarly determined by ELISA.

For antigen competition assays and Western Blot analysis of mAb specificity, we produced MDA-BSA samples with different levels of modification. Briefly, MDA was generated by acid hydrolysis of tetramethoxypropane (Sigma Aldrich), thereafter molecular grade BSA (NEB) at 1 mg/ml was modified by 50 mM MDA in PBS (pH 7.4) for 2-24 hrs at 37°C, followed by extensive dialysis to PBS. For generation of higher level of MAA-type modifications, acetaldehyde was added to 25 mM to the 2 hrs MDA-reaction. For ELISA competition studies, 100 ng/ml mAb IgG was mixed with indicated concentration of antigens in 1% BSA in PBS, incubated for 15 min at 37°C, and subsequently analyzed for binding to commercial MDA-BSA, as above. For SDS-PAGE and Western blot analysis, 3 µg of different lots of MDA-BSA or control-treated BSA were reduced and separated on Bolt Bis-Tris 4-12% gels with MES-SDS running buffer and blotted to PVDF membrane according to the manufacturer's instructions (ThermoFisher Scientific). The membrane was blocked with 3% BSA and stained with 1 µg/ml 1276:01F04 4°C o/n, followed by detection with biotinylated goat (Fab')2 anti-human IgG (Jackson ImmunoResearch) and anti-biotin-HRP (Cell Signaling Technology). Binding was detected by chemiluminescence using Clarity Western ECL Substrate (BioRad).

For ELISA studies of purified human fibrinogen (Sigma Aldrich), human serum albumin (HSA, Sigma Aldrich)), purified bovine histone 2B (Immunovision), or recombinant human vimentin (kind gift from Dr Karl Skriner, Charité Universitätsmedicin, Berlin), the antigens were coated at 3 ug/ml and MDA-modified on the surface for 2 hrs at 37°C by adding 100 mM MDA in PBS, followed by washing and blocking. Citrullination of fibrinogen and histone 2B was performed in 100 mM Tris, 10 mM $CaCl_2$, 5 mM DTT, with PAD4 (Cayman Chemicals, 0.75 U/mg protein), 37°C 2 hrs, followed by dialysis to PBS. Vimentin was similarly citrullinated using rabbit PAD (Sigma Aldrich).

All mAbs were also screened for binding to citrullinated peptide antigens using an antigen microarray solid-phase allergen chip multiplex assay (ISAC) with paired citrullinated and arginine-containing peptides, as well as other control antigens (Phadia AB, Uppsala, Sweden) [38]

(Supplemental Table 5, Supplemental Figure 9). Reactivity to the diagnostic peptide CCP3 was in addition tested for mAbs at 5 ug/ml with the QUANTA lite CCP3 assay (Inova Diagnostics).

## 2.5 Inflammatory biomarkers

Serum concentrations of inflammatory biomarkers and cytokines (IL-17F, IL-1β, TNF-α, IL-6Rα, IL-6, VEGF, sTNFRII and CRP) in the DMARD naïve cohort was determined using the highly sensitive Singulex Immunoassay System (Singulex, Inc., Alameda, CA) [39].

## 2.6 Osteoclast stimulation assay

Peripheral blood mononuclear cells (PBMCs) were isolated from healthy blood donor buffy coat by ficoll separation (Lymphoprep; Axis Shield, Norway) and monocytes were positively selected using anti-CD14 microbeads (Miltenyi Biotec). CD14-positive monocytes were seeded in 96-well plates 1x10^5 per well in 200 µl and differentiated into macrophages in Dulbecco's modified Eagle medium (DMEM) supplemented with 25 ng/mL macrophage colony-stimulation factor (M-CSF) (Peprotech). After every 3 days half of the medium replaced supplemented with 30 ng/mL M-CSF, 2 ng/mL RANKL (R&D Systems) along with 1 µg/ml or 10 µg/ml mAbs. The osteoclast (OC) culture was stopped after 8 days treatment. OCs were stained using tartrate-resistant acid phosphatase (TRAP) staining (leucocyte acid phosphatase kit 387A, Sigma-Aldrich) and analyzed. TRAP-positive cells with at least three nuclei were counted as OCs using a light microscope. In parallel, the OCs were generated in synthetic calcium phosphate surface (Corning) for 14 days and the surface area eroded was analyzed by the NIS-elements from Nikon (BergmanLabora).

## 2.7 Immunohistochemistry

Binding of anti-MDA mAbs to synovial tissue was evaluated by immunohistochemistry. In brief, 2% formaldehyde-fixed 7-µm-thick cryostat sections of synovial biopsy tissue was washed and permeabilize by PBS/saponin (0.1%, pH 7.4). Tissue sections were blocked with 1 % hydrogen peroxide 50 mins, followed by 30 min 3 % BSA 5 µg/ml human Fc-block (BD Bioscience), and stained with biotinylated (EZ-Link Sulfo-NHS-LC-Biotin, Thermo Fisher) human IgG1 clones at 2 µg/ml for 2 hrs at room temperature. Binding was detected with Vectastain elite ABC HRP kit (Vector Laboratories) and DAB Substrate kits (Vector Laboratories) and subsequently counterstained with Mayer's haematoxylin and viewed using a light microscope (Reichert Polyvar 2 type 302001, Leica). Results were evaluated by scoring the binding: 0, no staining; 1, low amount of staining; 2, intermediate staining and 3, high staining.

### 2.8 Statistical analysis

For statistical analysis, Prism (Graphpad) was used to assess for differences between groups and for correlations between measurements. Mann-Whitney test was used for comparing two groups and Spearman correlations were used for evaluation of correlation between measurements as indicated. P-values <0.05 were considered statistically significant.

### 2.8 Immunoglobulin gene analysis and structure modeling of antibody variable regions

The V-(D)-J genes of the immunoglobulin variable regions of isolated B cells were evaluated using the V-QUEST or IgBLAST web tools, where the closest germline and mutation rate was determined by comparison to the international Immunogenetics information (IMGT) database for human Immunoglobulin genes [40]. Models of the variable region encoded surfaces from the selected clones were generated with Prediction of Immunoglobulin Structure (PIGS) web server [41], using the best H and L chain method. Illustrations of the structures were generated in Jmol: an open source Java viewer for chemical structures in 3D (http://www.jmol.org/). Molecular electrostatic potential (MEP) calculations were calculated with the PDB2PQR server [42] and Jmol MEP surface using RWB color scheme (scale -0.5, 0.5).

## 3. Results

### 3.1 MDA-modified proteins are present in the joint of RA patients

Citrullinated proteins have previously been identified in RA synovial tissue using mass spectrometry [30] and in the current study we investigated whether MDA-modified proteins are also locally generated during pathogenesis. The same seven patient samples were re-analyzed, with identification of MDA-modified peptides in the synovial extracts by MS/MS analysis. After initial data-filtering and high accuracy validation, in total 29 MDA-peptides were identified from 10 different MDA-modified proteins (Table 1, Supplemental Table 1, Supplemental Figure 1). Examples of MDA-modification were found on all amino acids with amine or amide groups (i.e. lysine, asparagine, glutamine, histidine and arginine). A range of MDA-modified proteins were detected, which included high abundant serum proteins (i.e. albumin, hemoglobin, IgG gamma chain, and transferrin), but also the lipid metabolism associated apolipoprotein A1, the acute phase protein $\alpha$1-glycoprotein-1, and the enzyme carbonic anhydrase-1. We also identified MDA-modification of the cytoskeletal proteins actin and vimentin, in our RA samples. Intriguingly, two vimentin peptides (440-445 and 446-466) and an actin peptide (62-68) have previously been reported also undergo citrullination at the same arginine position that was now detected as MDA-modified [30]. In the seven different synovial samples, we observed large differences in the detected abundance of MDA-modification, ranging from high level to intermediate, and no detection. While the highest levels were seen in two ACPA seropositive RA patients, MDA-modification was also detected in the ACPA seronegative patient samples. While a number of synovial proteins were found to be MDA-modified in the RA joint including proteins that are also found to be citrullinated, the level of citrullination did not correlate with the level of MDA-modification, and the patients with highest MDA-detection were not the same patients with high levels of citrulline-modification [30].

### 3.2 IgG and IgM anti-MDA are increased in rheumatoid arthritis

To consider the relevance to autoimmune pathogenesis we investigated the reactivity of patient serum antibodies with MDA-modified self-proteins. Studies of 162 RA, 25 psoriatic arthritis (PsA), and 30 osteoarthritis (OA) patients, and 71 healthy controls, demonstrated that levels of IgG anti-MDA were significantly elevated only in patients with PsA (p=0.01) and RA (p<0.0001), compared to controls (Table 2, Figure 2). Similarly, IgM anti-MDA levels were significantly higher in PsA (p=0.006) and RA (p<0.0001). While healthy individuals had detectable levels of IgM anti-MDA antibodies, the levels of IgG anti-MDA were low or undetectable in these subjects.

Taken together, the immune dysregulation in RA patients was preferentially associated with elevated levels of autoantibodies to MDA-modified self-proteins, while levels of natural antibodies to other determinants (e.g. phosphorylcholine) may not show the same patterns.

### 3.3 IgG and IgM anti-MDA are elevated in new onset RA compared to established disease

In the initial analysis, we compared IgG anti-MDA responses in patients with new onset RA (NORA) disease, defined by less than six months' disease duration, to early RA (ERA), defined by disease duration between six months and two years, and to patients with chronic RA (CRA), with greater than two-year disease duration (Supplemental Figure 3). Particularly, we found the NORA patients had significantly higher levels of IgM anti-MDA (68±74 RU/ml vs 35±40 RU/ml, p=0.008) and IgG anti-MDA (21±13 RU/ml vs 10±8 RU/ml, p<0.0001), than CRA patients (Table 3). Significantly, the NORA patients had significantly greater disease activity than the CRA patients (DAS28 5.7±1.2 vs 4.8±1.5, p=0.01) and we could detect a direct correlation of IgG anti-MDA with DAS28 scores (p=0.0001, Spearman R=0.38).

To consider the potential effect of treatment, it is important to note that the NORA patients had not received any disease-modifying antirheumatic drugs (DMARDs), while the patients with established chronic disease were receiving standard of care DMARD treatment. For the 19 NORA patients, we evaluated levels of anti-MDA in serum samples obtained at two different time points, separated by 2-4 weeks, but there were no significant differences between these matched samples (Supplemental Figure 4). However, seven of these patients receive treatment with a synthetic DMARD (methotrexate, corticosteroids and/or plaquenil) between the first and second visit, and in these samples, the second time point displayed significantly reduced levels of IgG anti-MDA antibodies compared to the first time point, in paired analysis (p=0.04).

### 3.4 IgG anti-MDA levels are markers of inflammation in DMARD naïve patients

To directly explore the relationship between inflammation and the anti-MDA antibody response, we further studied a separate cross-sectional cohort of RA patients that had never received DMARD treatment (i.e. DMARD naïve) (Supplemental Figure 3, Table 4). In this cohort, there was also a significant albeit weak correlation for IgG anti-MDA with DAS28 (n=62, Spearman R=0.29, p=0.03). When comparing ERA and CRA DMARD naïve patients with similar disease activity by DAS28, there were no significant differences in IgG anti-MDA levels. Similarly, there were no differences in IgG anti-MDA between DMARD naïve CRA patients compared to CRA patients receiving standard of care DMARD treatment, in two groups with similar mean DAS28 scores (Supplemental Table 3, Supplemental Figure 5). In summary, the higher levels of IgG anti-MDA antibodies that were found in NORA compared to CRA patients may primarily reflect an association with higher disease activity. Yet, the ERA patients in the DMARD naïve cohort also had a longer disease duration (1-2 years) compared to the NORA patients (<6 months), so we cannot completely exclude effects that reflect the duration of symptomatic disease. However, we hypothesize that IgG anti-MDA correlates with elevated inflammation associated with active disease. We therefore proceeded to analyze inflammation-related biomarkers and measurements in the DMARD naïve patients.

From the analyses of 62 DMARD naïve patients, we found significant direct correlations between serum levels of IgG anti-MDA and the pro-inflammatory cytokines IL-17F, IL-6, and TNF-alpha (p=0.05, Spearman R=0.26; p=0.03, Spearman R=0.27; p=0.002, Spearman R=0.39, respectively; Table 5). Furthermore, higher levels of IgG anti-MDA also significantly correlated with raised erythrocyte sedimentation rates (ESR) and C-reactive protein levels (CRP) (p=0.002, Spearman R=0.39; p=0.003, Spearman R=0.37, respectively). However, the DAS28 measurement used in our studies incorporated ESR measurements, and only CRP can therefore be considered an independent variable. As IgG anti-MDA are increased in parallel to pro-inflammatory cytokines, we interpret our findings as evidence that levels of these antibodies are indeed a direct reflection of systemic inflammation.

### 3.5 Serum IgG anti-MDA levels correlate with clinical disease activity

Based on the above-described correlations between IgG anti-MDA and disease activity, we repeated these analyses in the complete RA cross-sectional cohort. Indeed, significantly elevated serum levels of IgG anti-MDA antibodies were observed in patients with moderate disease (DAS28 3.2-5.1) as well as in patients with high disease activity (DAS28>5.1), compared to patients with low disease or in DAS remission (DAS28<3.2) (Table 6, Figure 4). A significant direct correlation was observed between IgG anti-MDA levels and DAS28 score (p<0.0001, Spearman R=0.36). There was no correlation between DAS28 and IgG anti-CCP3 levels (Supplemental Figure 6). Indeed, IgG anti-MDA levels did not significantly correlate with serum IgG anti-CCP3, although IgG anti-MDA generally appeared to be higher expressed in a subset of anti-CCP- and/or RF-seropositive patients (Supplemental Figure 6). Although there were weak correlations between IgA anti-MDA and IgA anti-CCP3 levels (R=0.29, p=0.02) and similarly between IgM anti-MDA and IgM anti-CCP3 levels (R=0.30, p=0.008).

To evaluate IgG anti-MDA reactivity as a potential disease activity biomarker, we found the odds ratio for high DAS28 (>5.1) for a positive anti-MDA IgG test (with a cutoff of 15 RU/ml) was 2.4 (Fisher's exact test CI:1.2-5.1, p=0.02) with sensitivity of 80%, although the specificity only reached 36%. These associations were explained by the fact that only a subset of 28% of RA patients in this cross-sectional cohort had significantly elevated IgG anti-MDA autoantibodies. We speculate that the associations between elevated anti-MDA autoreactivity with more active disease may reflect a higher level of inflammation and oxidative stress. In turn, this increased burden of MDA modified proteins, generated either locally or systemically, could contribute to the triggering of anti-MDA immune responses. On the other hand, these correlations with elevated serum levels of autoantibodies do not necessarily reflect an active causal role of MDA-reactive B cells in RA pathogenesis.

### 3.6 Anti-MDA reactive B cells are prevalent in the RA joint

To directly investigate whether MDA-specific responses are involved in the pathogenesis of rheumatoid synovitis, we isolated single B cells reactive with MDA-modified proteins from synovial infiltrates from active RA patients. From the synovial B cells of eight ACPA seropositive and two ACPA seronegative RA patients (Supplemental Table 4), the antibody gene rearrangements were amplified, cloned, and then expressed as recombinant IgG1 monoclonal antibodies (mAbs) that were subsequently screened for MDA-reactivity by ELISA. Among 114 memory B cell- and 86 plasma cell-derived mAbs, significant MDA-reactivity was detected in six clones; four memory cell-derived and two plasma cell-derived mAbs (Figure 5; Supplemental Figure 8, Table 7). All of the MDA-reactive antibodies from memory B cells were isolated from ACPA seropositive patients. However, intriguingly, the antibodies from plasma cells with the highest MDA-reactivity were isolated from B cells from a seronegative RA patient.

These anti-MDA mAbs were then evaluated for binding reactivity with other types of non-enzymatic posttranslational modifications by ELISA (Figure 5-6, Supplemental Figure 8). Two out of four memory B cell-derived mAb clones demonstrated weak binding with carbamylated BSA, while one of the two plasma cell-derived clones showed polyreactivity with a range of control antigens. These mAbs were also extensively evaluated for possible citrulline-reactivity with a multiplex antigen microarray containing citrullinated peptides, as previously described [38], as well as by CCP3 ELISA test. The MDA-reactive clones neither showed significant binding reactivity with citrulline-containing ligands (Supplemental Figure 9), nor did the previously identified ACPA mAbs cross-react with MDA modified ligands [33](Figure 5-6). Similarly, the MDA-reactive clones also did not display detectable RF activity, as measured by rabbit IgG binding ELISA. Conversely, the strong RF-expressing B-cell clones displayed little or no reactivity with MDA-modified ligands by ELISA (Figure 5, Supplemental Figure 10).

In summary, three antibody clones (146+:01G07, 1276:01F04 and 1362:03H05) were found to be highly specific for MDA-modified epitopes and were devoid of cross-reactivity with citrullinated, carbamylated, 4-HNE-modifed or phosphorylcholine-containing antigens. These antibody clones were derived from different synovial B cell subsets; with one originating from an FcRL4+ IgA+ clone (146+:01G07), one from an IgG1+ memory B cell (1276:01F04), and the other from an IgG1-secreting plasma cell (1362:03H05). Importantly, the antibody clones recognized MDA-adducts independent of the protein context, and also displayed a specific binding reactivity with MDA-modified human albumin, MDA-modified fibrinogen, MDA-modified histone 2B and MDA-modified vimentin, in addition to MDA-modified bovine albumin (Figure 6). While the 1276:01F04 showed consistently strong reactivity to all tested MDA-modified antigens, 146+:01G07 (Figure 6) and 1362:03H05 (data not shown) had weaker binding and possibly a more preferential binding specificity for albumin and vimentin, although this could reflect differences in saturation level between these

mAbs. Similarly, these monoclonal antibodies, which displayed strong binding to MDA-oxidized human low density lipoprotein (MDA-LDL), had no binding reactivity with native LDL (Supplemental Figure 11). Collectively, these novel findings suggest a high representation of MDA-reactivity among synovial memory B cells (3.5%, 4/114) and synovial antibody secreting cells (2.3 %, 2/84), and highly specific anti-MDA autoreactive B cells therefore appear to be prevalent at the primary site of disease in this autoimmune disease.

### 3.7 Strong MDA-binding in close-to-germline encoded synovial antibody

The antibody gene nucleotide sequences of the isolated synovial anti-MDA mAbs were analyzed in comparison to their closest V-(D)-J germline sequence in the IMGT database, and if they had more than two nucleotide mismatches in either the heavy or light chain they were determined to have somatic hypermutations. Interestingly, the B cell clone, 1276:01F04, which was isolated from an IgG1-bearing synovial memory B cell, displayed the strongest level of reactivity in all MDA-binding assays, was encoded by close-to-germline configuration variable gene rearrangements (≤2 mismatches in heavy and light chains: VH4-39*01 had one replacement mutation; VL1-51 had two silent mutations, Table 7). Furthermore, in the light chain variable region, these two silent mutations are near the 5' end of the variable region gene and are likely to have arisen from the PCR primer used for the cloning method. In the heavy chain variable gene, the single nucleotide mutation from the closest known germline gene encodes for an amino acid A to V replacement mutation in the third complement determining region, CDR3, (HCDR3: ARVRGYFDY => VRVRGYFDY), which could represent unknown allelic variation, the effect of untemplated N addition(s), or possibly an error introduced by PCR. While this could reflect somatic hypermutation, this conservative change may not in any case affect the specificity or strength of the binding interaction. Nevertheless, in light of these very minor sequence variations we speculate that this antibody is derived from the natural antibody pool. Our findings are also consistent with an earlier report that MDA/MAA-reactive B cells in human umbilical cord blood are enriched in the representation of VH4-39*01 rearrangements [26]. Yet, all the other isolated synovial anti-MDA clones displayed changes consistent with extensive somatic hypermutation, with 6-18 mutations in either the light chain or heavy chain, which may suggest they had undergone affinity maturation.

The 1276:01F04 mAb exhibited strong MDA-ELISA reactivity even at low concentration (<50 ng/ml IgG1), and the binding was specifically inhibited by soluble MDA-modified protein in competition experiments (Figure 6). The capacity for inhibition of binding was modulated by the degree of MDA-modifications, whereas BSA subject to 24 hr modification was a stronger inhibitor than MDA-BSA generated by 2 hr treatment. Similarly, addition of acetaldehyde to the MDA reaction (i.e. MAA) generated an antigen with stronger interaction with 1276:01F04. MS analysis reveals that the *in vitro* MDA-modified proteins without addition of acetaldehyde, still display chemical structures associated MAA (i.e. MDHDC) that increased with higher *in vitro* modification times, presumably due to spontaneous breakdown of the excess MDA to acetaldehyde, along with subsequent MAA reactions (Supplemental Table 2, Supplemental Figure 1 and 12). Hence, this mAb may preferentially bind to MAA-type structures although our studies cannot absolutely rule out binding to the major MDA-adduct N-ε-(2-propenal)-lysine. SDS-PAGE and Western blot analysis of different MDA-modified protein batches revealed a low but detectable level of aggregation and possible intermolecular cross-linking in our in-house generated lots. However, 1276:01F04 showed specific binding to all MDA-modifications and especially strong binding to the monomeric 24 hr treated MDA-BSA fraction (Supplemental Figure 12).

With molecular modeling tools, we studied the predicted structure of the antibody variable regions. 1276:01F04 demonstrated a central groove-like structure formed by the surface generated by the heavy and light chain CDR-loops (Figure 7). While it is not possible to know the true antigen-binding surface without the solved structure of the antibody-antigen complex, based on the model we speculate that the antibody may also have the capacity to recognize larger structures, and in addition to these small molecule modifications that are about the size of an amino acid side chain. By contrast, modeling of the mAb clone 146+:01G07 revealed a deep pit-like deep structure, which may suggest a smaller potential binding-surface. While high local hydrophobicity or long extended loops or have been implicated in antigen-binding sites with polyreactivity [43], 1276:01F04 instead has a rather short HCDR3 and is devoid of such structural features.

In general, all the anti-MDA monoclonal antibodies displayed substantial solvent exposed positively charged surfaces, with 146+:01G07 and 1362:03H05 having more charge than 1276:01F04. This was also reflected in somewhat higher estimated isoelectric point for the variable region (VL-VH) of 146+:01G07 and 1362:03H05 (9.3 and 9.1) compared to 1276:01F04 (8.8). However, despite the charged surface, these mAbs did not display polyreactivity to the prototypic polyanionic macromolecule, native dsDNA (data not shown). In general, these B cell clones with high specificity for MDA/MAA modifications, which were recovered from the RA synovium, display features that are commonly associated with the natural antibody pool.

### 3.8 Differential tissue binding activity of synovial B cell-derived mAbs

To investigate the relevance of the MDA reactive antibodies to RA pathogenesis, we performed immunohistological staining studies with human synovial tissue. Both the close-to-germline clone 1276:01F04 and the hypermutated monoclonal 146+:01G07 recognized antigenic targets present in the RA inflamed joint (Figure 8). However, their staining patterns were different, as regions in some tissue samples were strongly reactive with 1276:01F04 while others were more strongly reactive with 146+:01G07. Nevertheless, the clone 146+:01G07 had generally higher reactivity with these sections despite our above-described findings of lower reactivity with *in vitro* MDA-modified proteins. We also evaluated for staining of healthy synovial tissue, with tissue from RA patients or disease controls, including patients with spondyloarthropathy, osteoarthritis, and psoriatic arthritis. As a control, we compared the staining activity of synovial monoclonal antibody devoid of MDA- or ACPA activity. While this control monoclonal antibody also showed moderate binding to patient materials, the two anti-MDA mAbs generally had higher binding to all three synovial types. These findings are consistent with the *in vivo* generation of a range of MDA-related antigens during synovial pathogenesis, with variations in their recognition by different mAbs representative of the immune response.

### 3.9 Effects of anti-MDA mAbs on in vitro osteoclastogenesis

To evaluate the potential pro-inflammatory properties of anti-MDA antibodies, we studied the functional effect of anti-MDA mAb exposure in monocyte-derived human osteoclast cultures. In these assays, osteoclasts were generated from CD14-positive blood monocytes by stimulation with M-CSF and sub-optimal concentrations of RANKL, with or without the presence of IgG mAbs. We observed that the two hypermutated anti-MDA clones 1362:03H05 and 146+:01G07 robustly enhanced osteoclastogenesis. A significant increase in both the number of TRAP-positive multi-nucleated cells, as well as the level of bone resorption on synthetic calcium phosphate surfaces, was demonstrated with the mAbs compared to control conditions (Figure 8, Supplemental Figure 13). In contrast, the clone 1276:01F04 had no detectable effect on osteoclast differentiation. Taken together, these functional assays suggest that only the hypermutated anti-MDA IgG clones enhance osteoclastogenesis in assays designed to evaluate properties of antibodies that may contribute *in vivo* to joint destruction.

## 4. Discussion

Our studies reveal that MDA carbonylation-modified proteins are present in the RA synovium during active disease, and that RA patients have elevated levels of autoantibodies to these oxidation-associated protein modifications, especially at onset of disease, and that anti-MDA levels correlate with disease activity. Furthermore, a high frequency of MDA-specific B cells was found in the synovium of RA patients with active disease, and our data suggests that some anti-MDA IgG, but not others, may contribute to pathogenesis by enhancing the generation and activity of osteoclasts. We hypothesize that anti-MDA B cells may be selected within the joint due to interactions with the high local increased burden of ROS and MDA-modified proteins that arise in this disease milieu.

MDA-modification can lead to intra- and inter-molecular cross-linking of proteins and change their physical and functional properties. IgG antibodies to these modified proteins may have pro-inflammatory properties due to their effector functions and enhancing danger-signal of MDA/MAA, while IgM antibodies may be more likely to facilitate blocking and removal of potentially harmfully altered proteins. Free MDA can also react with acetaldehyde and together form MAA-type of modification of proteins. This has been observed to occur in alcoholic liver disease, cardiovascular disease, in cigarette smoking, and has also recently been shown in staining of RA synovial tissue [44-47]. MAA-adducts can contribute to airway inflammation and stimulate IL-8 production, and MDA-adducts have been shown exaggerate autoimmunity and IL-17 cell activation in murine models [48, 49]. Hence, MDA- and MAA-adducts may have pro-inflammatory properties in the context of RA. It is important to emphasize that, even though citrullination, being an enzymatic posttranslational modification, is regulated distinctly differently than ROS-associated modifications, for example neutrophils have the capacity to mediate both types of modifications. In fact, the pathway for generation of neutrophil extracellular traps (NETs), a process that has been implicated in the RA pathogenesis, involves both ROS release and PAD4-mediated citrullination [50]. Hence, in a setting of chronic inflammation, citrullination and MDA/MAA-modification could simultaneously arise. This was confirmed by the presented mass spectrometry analysis of RA synovial tissue. We speculate that the two different types of modifications could also occur within the same protein molecule. Hence, MDA-positive B cells could in theory act as potent antigen-presenting cell which also facilitate epitope spreading and activation of anti-citrulline reactive T-cells. The proportion of B cells with natural autoreactivity for oxidation-associated modification in the circulation are presumably highly represented due to a positive selection pressure rather than negative for these innate-like B cell populations (i.e. B-1 and marginal zone B cells), as demonstrated for other natural autoreactivites in murine models studying [51]. The threshold for selection and expansion of IgG anti-MDA/MAA B cells may therefore be lower than for disease-specific IgG to other autoreactivity-associated modifications, such as citrullination.

In the current studies, we have focused on MDA-reactivity rather than MAA-adduct reactivity since a lower degree of modifications may occur more frequently under (patho)physiological conditions. However, antibody reactivity to MDA-adducts, compared to MAA-adducts, can be hard to distinguish with absolute confidence. A more extensive cross-sectional study demonstrated an increase of anti-MAA reactivity in RA, which associated with ACPA positivity without any direct cross-reactivity between the MAA and citrulline distinct epitope bindings [52]. These important data have opened up a discussion about oxidization-associated antigens in RA [53], and the results are in-line with our observation that synovial anti-MDA antibodies can have high specificity for MDA-modifications whilst showing no reactivity to other oxidation associated carbonylation, such as 4-hydroxynonenal (4-HNE), nor to carbamylated or citrullinated proteins.

Importantly, we demonstrate that anti-MDA antibodies bind specifically not only to the surrogate antigen MDA-BSA, but also to other more RA-relevant modified proteins such as MDA-histone, MDA-fibrinogen and MDA-vimentin. Other studies have suggested that certain murine or human natural antibody clones that bind MDA modified low-density lipoprotein (LDL) have the ability to cross-react to *Porphyromonas gingivalis* Gingipain protein, or to carbamylated epitopes [54, 55]. The reported phage display selected human carbamyl/MDA cross-reactive clones had some similarities with the 146+:01G07 with unmuted VH3-33/3-30 gene usage although 146+:01G07 showed no carbamylated-protein reactivity. Similarly, the anti-MDA synovial B cell clone 1276:01F04 shows similarity in VH gene usage with previously reported cord blood anti-MAA natural antibodies that had a high VH4-39*01 bias [26]. Natural IgM immunity to oxidation-associated MDA adducts, especially in oxLDL, has been extensively studied for their protective properties in atherosclerosis [18, 26] (reviewed in [13, 27]), but their relation to inflammation-associated IgG anti-MDA is not well known. Natural antibodies have been defined as primarily spontaneously secreted IgM (and IgA), part of T-cell independent responses, and are considered to be primarily produced by innate-like B-1 cells with limited BCR repertoires and (near) germline encoded variable regions [56]. Yet a recent study suggested that anti-MDA antibody production can, at least to some extent, be T-cell dependent [57], and therefore may not be solely produced by innate-like B cells. Although B-1 cells are also able to interact with T cells as antigen presenting cells [58], and human B-1 cells may also at time undergo isotype class switch [59].

Intriguingly, the synovial anti-MDA B cell clone 1276:01F04 had significantly stronger binding than any other clone, and was generated from an IgG1-bearing memory B cell. The high reactivity yet close-to-germline encoded variable regions is also consistent with the notion that this antibody is from a natural antibody producing B cell. Importantly, this particular antibody clone was devoid of pro-inflammatory properties in the *in vitro* osteoclast assay. In contrast, our other two MDA-specific clones, 146+:01G07 and 1362:03H05, had somatic hypermutations (in the case of 146+:01G07 only in the light chain), and displayed different binding patterns compared to 1276:01F04, with lower reactivity and possibly a preference for modified albumin and vimentin. Both types of antibodies showed binding to human synovial tissue samples, although the binding patterns were somewhat divergent and the 146+:01G07 clone generally had higher reactivity. We postulate that these represent a distinctly different class of anti-MDA reactivity, and future studies are needed to further characterize the molecular basis for antigen recognition beyond what we were able to detect in our limited assays.

Recent studies have demonstrated that anti-citrulline IgG may have previously unsuspected direct functions in the pathogenesis of arthritis by mediating bone destruction and pain induction [60-62]. One of the key events in this process is ACPA-mediated stimulation of osteoclastogenesis and IL-8 production [61, 62]. We speculate that some of these same properties may also be associated with pathogenic anti-MDA antibodies. Importantly, our results show that, certain hypermutated low-reactivity type, but not the natural antibody high-reactivity type, anti-MDA modified protein antibodies have the ability to significantly enhance osteoclast differentiation. Interestingly, osteoclast stimulating properties were seen in the 146+:01G07 mAb that originates from a FcRL4+ RANKL overexpressing, potentially pathogenic, B cell subset [31]. The level of this stimulation was similar to what has previously been observed for polyclonal anti-CCP2 ACPA. Importantly, we do not see neither anti-MDA reactivity in the polyclonal anti-CCP2 pool nor any citrullinated peptide reactivity of the anti-MDA antibodies (Supplemental Figure 13) suggesting that these are parallel reactivities.

These antibodies may have a more preferential reactivity to MDA-vimentin than other antigens and low but significant cross-reactivity with full-length citrullinated and native vimentin protein. Vimentin has been proposed as a potential pathogenic target recognized by ACPA on osteoclast [60] and is a well validated object of synovial citrullination [30, 63]. We can now also verify that vimentin is MDA-modified in the joint. It is intriguing that osteoclast differentiation is dependent on both mitochondrial ROS activation and PAD citrullination [61, 64] and that this may result in parallel protein modifications, alteration of protein functions, and pro-inflammatory antigen exposure. In conclusion, the similarities and differences between ACPA and anti-MDA induced osteoclastogenesis, as well as the pathogenic differences between IgG clones within the groups, merits further more extensive investigation. However, we speculate that it may be possible that IgG autoantibody-targeting the same antigen by different epitope-interactions and subsequently engaging activating Fc receptors or other stimulatory pathways could provoke the same downstream pathogenic response. Appreciation of the contribution of anti-MDA responses to RA pathogenesis may provide an additional reasons for the development of therapeutic regiments that restores physiologic immune tolerance [65]. Still, it is important to acknowledge that elevated anti-MDA autoreactivity is not unique to RA but also seen in other autoimmune diseases. An increase in anti-MDA in chronic inflammation may reflect higher levels of oxidative stress in general. This is in concordance with the variation in IgG anti-MDA levels with disease activity. Furthermore, not all natural antibody reactivities showed the same association. In the clinical screenings, we also evaluated antibody reactivity with phosphorylcholine (PC), an oxidation-associated lipid head group epitope, for which IgM antibody levels are reported to directly correlate with atheroprotection [28, 37, 66-68]. Both IgM and IgG anti-PC antibody levels were significantly higher in RA patients compared to controls. However, in contrast to our findings for anti-MDA responses, the healthy controls had high relative representation of IgG anti-PC antibodies, and no correlation with disease activity or pro-inflammatory markers were seen in the RA patients. Monitoring disease activity is critical for assessing the efficiency of therapeutic options and improving patient care. The 28-joint disease activity score, DAS28, which incorporates tender and swollen joint counts, the physician's global assessment, and an inflammatory marker, (i.e, ESR or CRP) [69], has become widely accepted and for many practitioners is the gold standard. However, there is a complete absence of consideration of the contribution of the dysregulated adaptive immune compartment from which this autoimmune disease arises. Future investigations of longitudinal cohorts are needed to better understand the clinical significance of these responses and to potentially refine a more practical and relevant biomarker.

### Concluding remarks

Much is still not known about chronic inflammation, oxidative stress, and how autoantibodies and autoreactive B cells mediate pathogenesis. Autoreactivity to oxidation-associated determinants and certain post-translational protein modifications are abundant in health, yet these are also clearly elevated in disease. Since anti-MDA immunity is part of the natural antibody responses, tolerance may be more readily breached in a setting of generalized disease associated inflammation and increased immune activation. Even though anti-MDA responses are not specific to patients with RA, these antibodies may contribute to the disease process through immune-complex mediated and FcR-engaging pathways involving more classical types of polyclonal IgG autoimmune responses. Better understanding of the complexity of the adaptive autoimmunity in RA may provide new insights in the pathogenesis, which will help to develop improved clinical disease monitoring tools and more specific and effective therapeutic strategies targeting the pathogenic autoreactivity.

### Acknowledgements
We thank Lena Israelsson, Dr Monika Hansson and Ragnhild Stålesen (Karolinska Institutet, Stockholm) for excellent support in antibody production, validation and characterization. Dr Karl Skriner and Peter Sahlström (Charite Universitätsmedizin, Berlin) for kindly providing recombinant native and citrullinated vimentin, Dr José Scher (New York University, New York, NY) for contribution of clinical serum samples, the Karolinska Hospital orthopedic surgery team for synovial tissue materials and Dr Erik af Klint (Karolinska University Hospital, Stockholm) for synovial biopsies from healthy volunteers.

This work was supported by the Swedish Research Council (2013-03624), Åke Wiberg's foundation (M15-0087, M16-0060), the Swedish Rheumatism Association (R-562111; R-660871), Nanna Svartz foundation (2015-00077), Ulla and Gustaf af Ugglas foundation (2016-00351) and King Gustaf V's 80-year foundation (FAI-2014-0005).

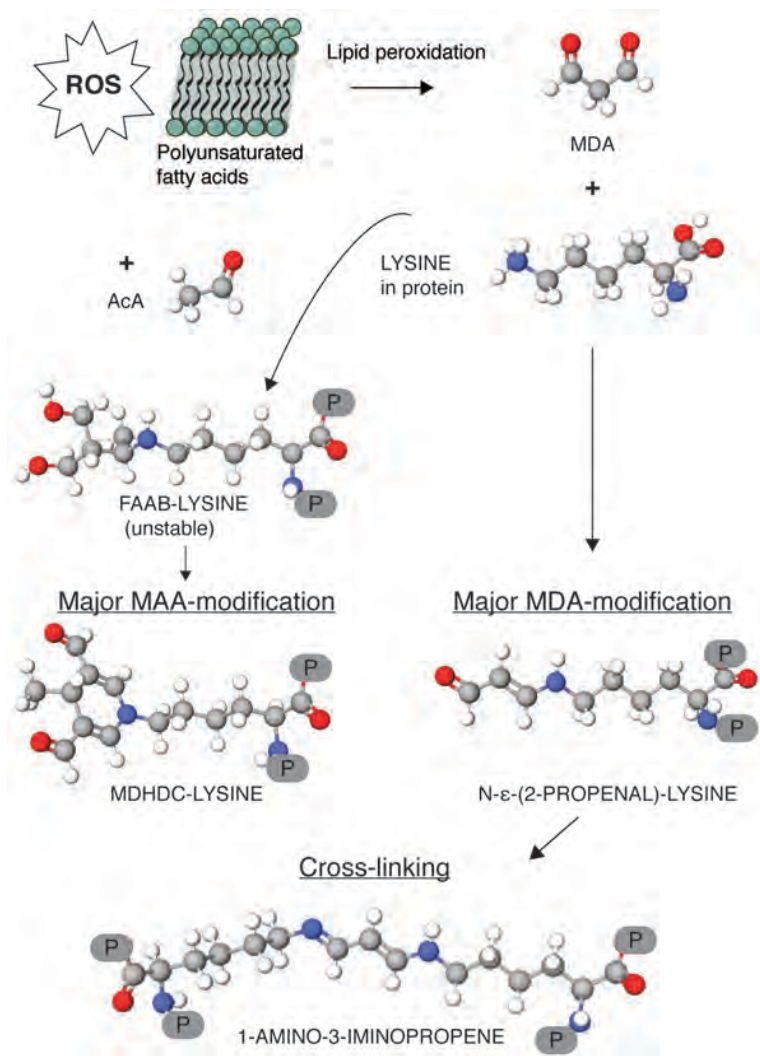

**Figure 1. Schematic overview of MDA and MAA adducts**

Malondialdehyde (MDA) can be produced during lipid peroxidation where excess reactive oxygen species (ROS) react with polyunsaturated fatty acids in cell membranes or lipid containing biomolecules. MDA can covalently modify protein residues with free amine groups (i.e. lysine, arginine, histidine and the N-terminus) and to lesser extent residues with amide groups (asparagine and glutamine). Here modification of lysine, the most commonly MDA-modified amino acid, is presented. MDA-modification of lysine generates N-$\varepsilon$-(2-propenal)lysine groups and can also lead to inter- or intra-molecular crosslinking and 1-amino-3-iminopropene adducts. More complex MAA structures are formed if acetaldehyde (AcA) is present in addition to MDA (in 1:2 ratio), leading to unstable 2-formyl-3-(alkylamino) butanal (FAAB) adducts or stable fluorescent 4-methyl-1,4-dihydropyridine-3,5 dicarbaldehyde (MDHDC) adducts, also denoted DHP-lysine residues. Grey, carbon; White, hydrogen; Blue, nitrogen; Red, oxygen. P, protein backbone.

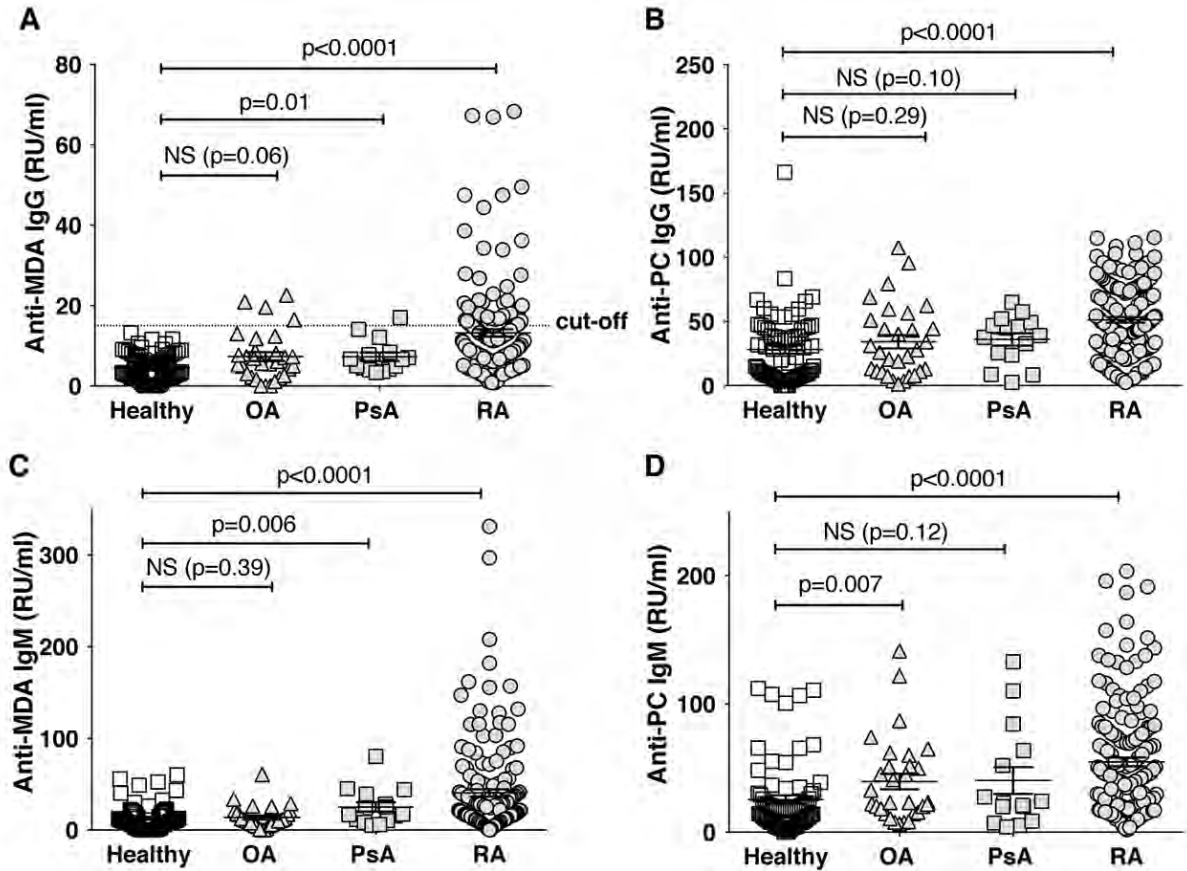

**Figure 2. Anti-MDA autoreactivity is increased in autoimmune disease**

ELISA screening of oxidation-associated autoantibody levels in 71 heathy blood donors, 30 osteoarthritis (OA), 25 psoriatic arthritis (PsA), and 162 rheumatoid arthritis (RA) patients, showed significantly increased levels of IgG and IgM anti-malondialdehyde (MDA) modified protein antibodies in autoimmune disease. **A, C**. Anti-MDA modified protein IgG and IgM, respectively. **B, D.** Anti-phosphorylcholine (PC) IgG and IgM, respectively. MDA-BSA and PC-BSA were used as screening antigens. P-values were derived from Mann-Whitney analysis.

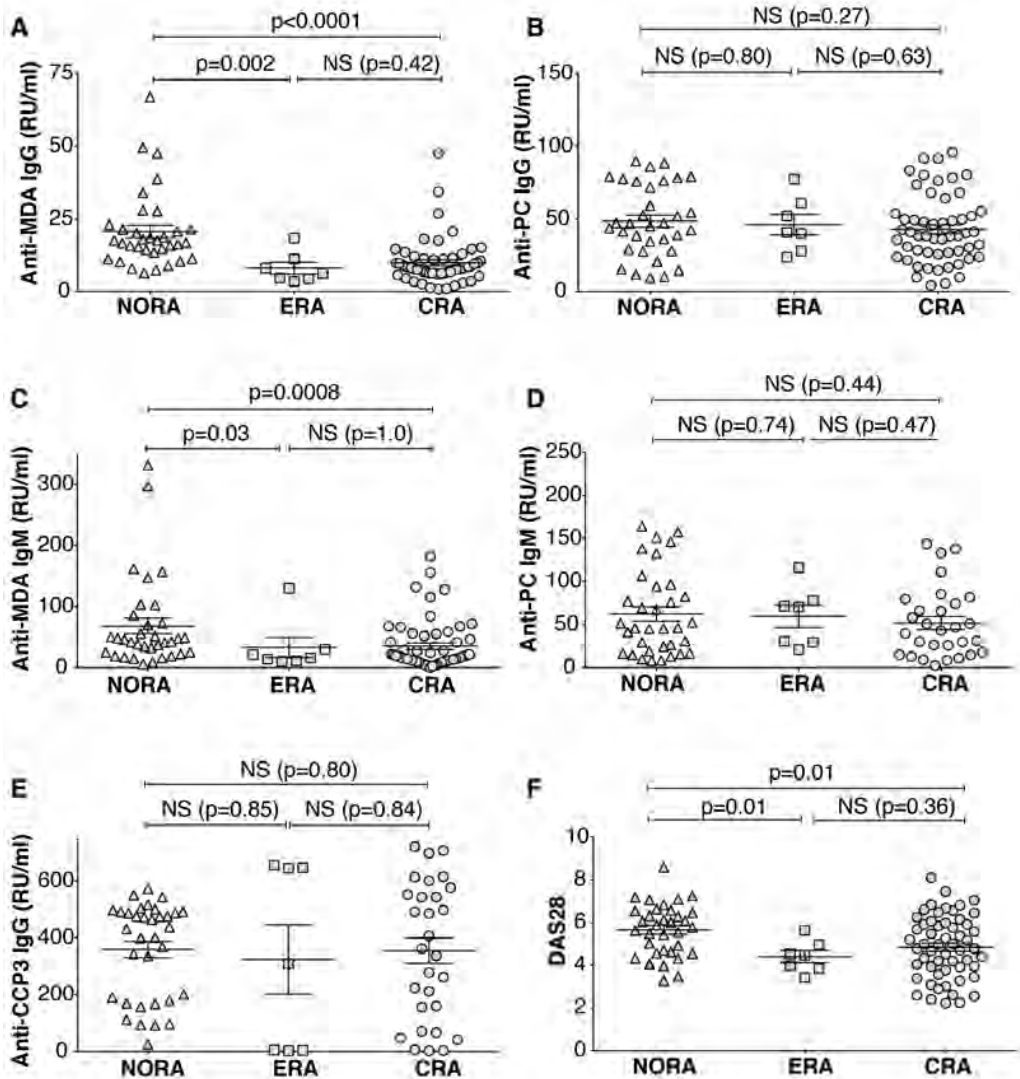

**Figure 3. Anti-MDA autoantibody levels are elevated in new onset RA patients compared to chronic patients**

Comparison of antibody levels by ELISA serum screening in RA patients with new onset disease (NORA, n=34) DMARD naïve patients with less than six-month disease duration, early RA patients (ERA, n=7) with more than six months but less than two years of disease duration, and chronic RA patients with more than two-years disease duration (CRA, n=56), demonstrates higher IgG anti-MDA-modified protein levels in NORA patients but also significantly higher disease activity. **A.** IgG anti-MDA, **B.** IgG anti-PC, **C.** IgM anti-MDA, **D.** IgM anti-PC, **E.** IgG anti-CCP3 **F.** Disease activity by DAS28 ESR. P-values are presented from Mann-Whitney analysis.

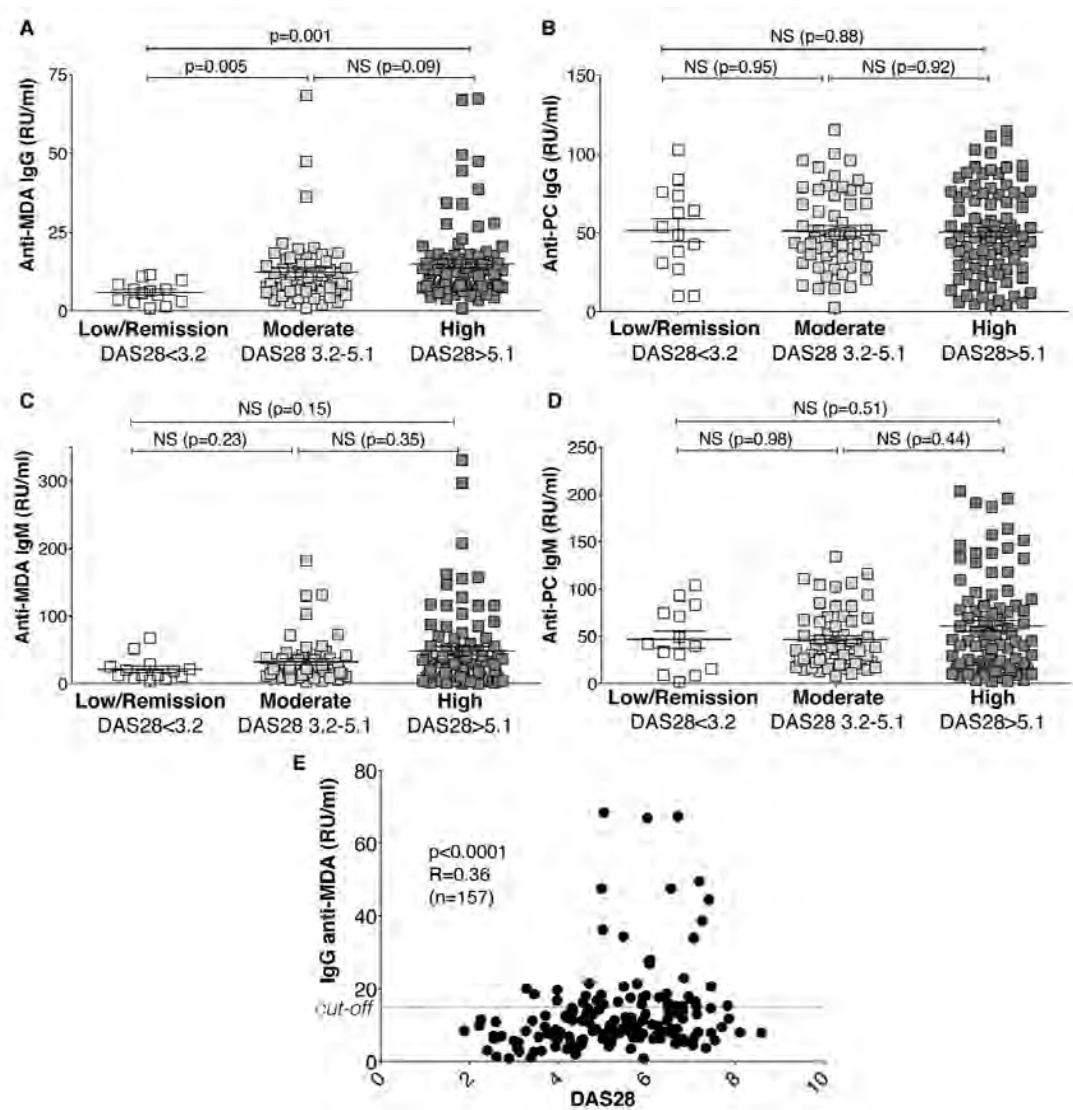

**Figure 4. Autoantibodies to MDA-modified proteins are significantly increased in patients with active disease**

Analysis of serum antibody levels by ELISA in RA patients with low disease activity (DAS28 ESR< 3.2, n=14), moderate disease activity (DAS38 ESR 3.2-5.1, n=56) or high disease activity (DAS28 ESR>5.1, n=87), showed significantly higher levels of anti-MDA in patients with high disease activity. **A.** IgG anti-MDA, **B.** IgG anti-PC, **C.** IgM anti-MDA, **D.** IgM anti-PC. P-values presented from Mann-Whitney analysis. **E.** Spearman correlation between DAS28 ESR and IgG anti-MDA levels. Among the patients with low disease, seven patients were considered to be in remission with DAS28<2.6.

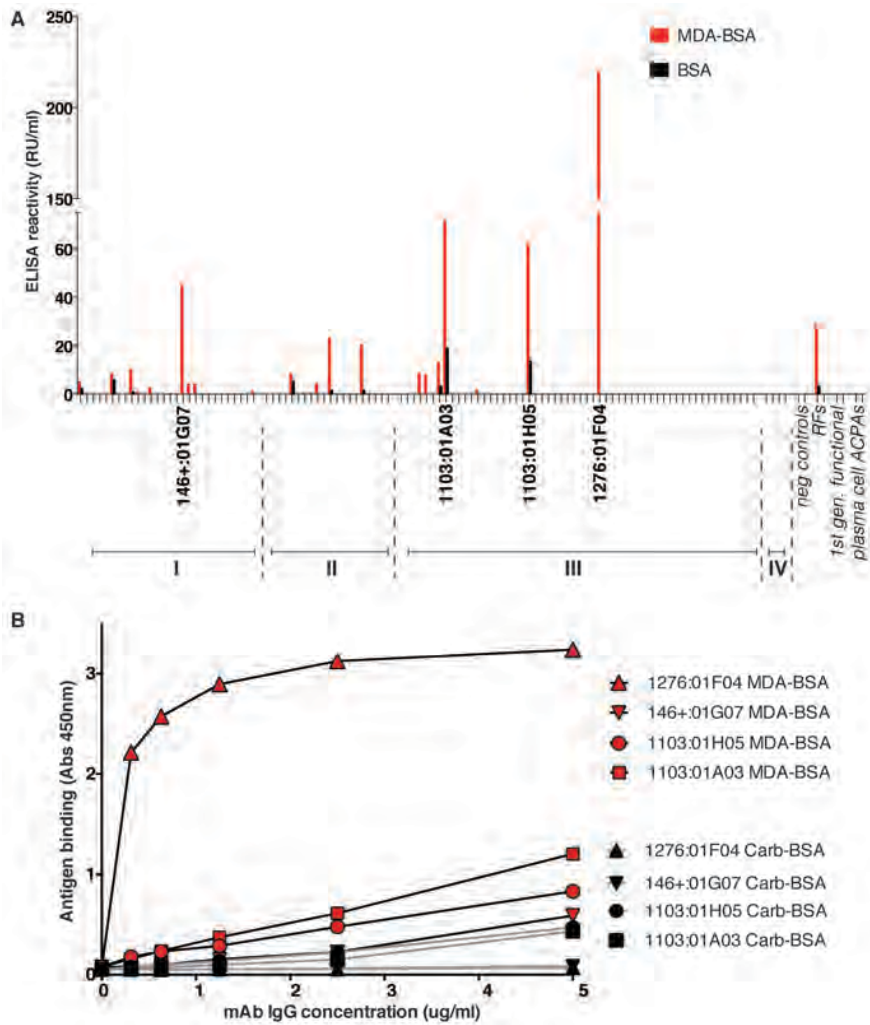

**Figure 5. MDA-reactivity in synovial memory B cells**

ELISA screening of purified recombinant human monoclonal antibodies isolated by single cell sorting of RA synovial memory B cells, cloned, and expressed in Expi293 cells as human IgG1. **A.** 114 memory B cell mAbs were screened at 5 ug/ml for binding to MDA-BSA and to control wells coated with unmodified BSA. Reactivity was quantified as Relative Units (RU)/ml based on a standard reference curve. Antibodies were derived from different synovial subsets: *I.* CD19+ FcRL4+ cells (n=28), *II.* CD19+ FcRL4- cells (n=20), *III.* CD19+ IgG + cells from CCP positive patients (n=56) *IV.* CD19+ IgG+ cells from CCP negative patients (n=4). On the right side of the panel values for control mAbs (n=8) are shown with two negative controls (1276:01G09, 1362:01E02) [32], two mAbs with RF activity (146-:01B05, 1276:01C11) [31, 32], two first generation pathogenic functional mAbs from memory cells (1276:01D10, 1103:01B02) [32] and two plasma cell derived ACPA mAbs (1325:01B09, 1325:04C03). [33] **B.** Four mAbs showed significant MDA reactivity (146+:01G07; 1103:01A03, 1103:01H05, 1276:01F04) and were further evaluated for MDA-BSA binding in serial dilution compared to reactivity to the control antigen carbamylated BSA (Carb-BSA).

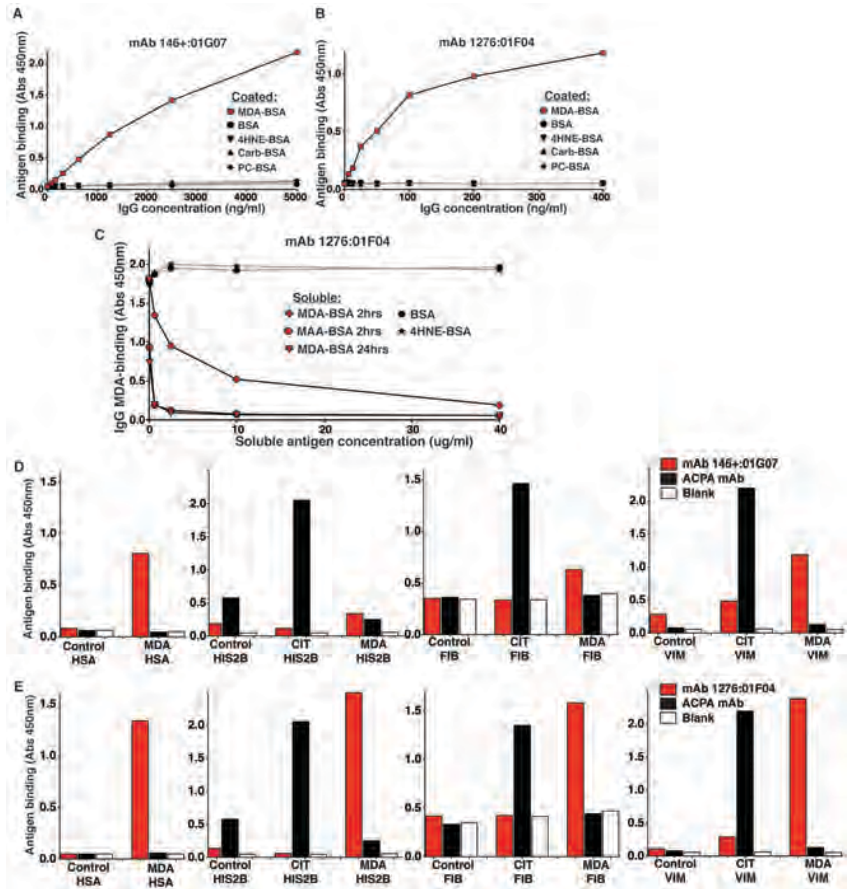

**Figure 6. High specificity of MDA-reactive synovial memory B cell derived mAbs**

The memory B cell derived recombinant antibody clones 146+:01G07 and 1276:01F04 demonstrated high specificity for MDA modification and did not bind to 4-HNE modified, carbamylated or PC-conjugated BSA. **A** and **B** shows serial dilution of mAbs 146+:01G07 and 1276:01F04, respectively, to different antigen surfaces coated on the same ELISA plate. **C.** Competition assay with soluble antigens. The mAb 1276:01F04 concentration was kept constant at 100 ng/ml and analyzed for binding to wells coated with commercial MDA-BSA at 3 ug/ml (Academy Bio-Medical) after 15 min 37°C pre-incubation with soluble in-house prepared MDA modified BSA with different degree of modification at indicated concentrations. The binding could be specifically blocked with MDA-BSA and not with control antigens (native BSA or 4-HNE modified BSA) and higher degree of MDA modification (24 hrs prepared MDA-BSA or MAA-BSA) showed higher inhibition capacity. **D and E** show binding to MDA-modified human albumin (MDA-HSA) MDA-modified human fibrinogen (MDA-Fib), MDA-modified bovine histone 2B (MDA-HIS2B) and human recombinant vimentin (MDA-VIM) compared to citrullinated proteins (PAD4 citrullinated fibrinogen and histone 2B, rabbit PAD citrullinated vimentin) or control treated proteins. The ACPA mAb 1325:01B09 with a preference for citrullinated fibrinogen and histone was used as a control for citrullination for fibrinogen and histone 2B and the ACPA mAb 1325:04C03 with a preference for citrullinated vimentin as control in the vimentin assay. MDA-modification was performed on the plate with 100 mM MDA for 2 hrs at 37°C. All mAbs were tested at 5 ug/ml.

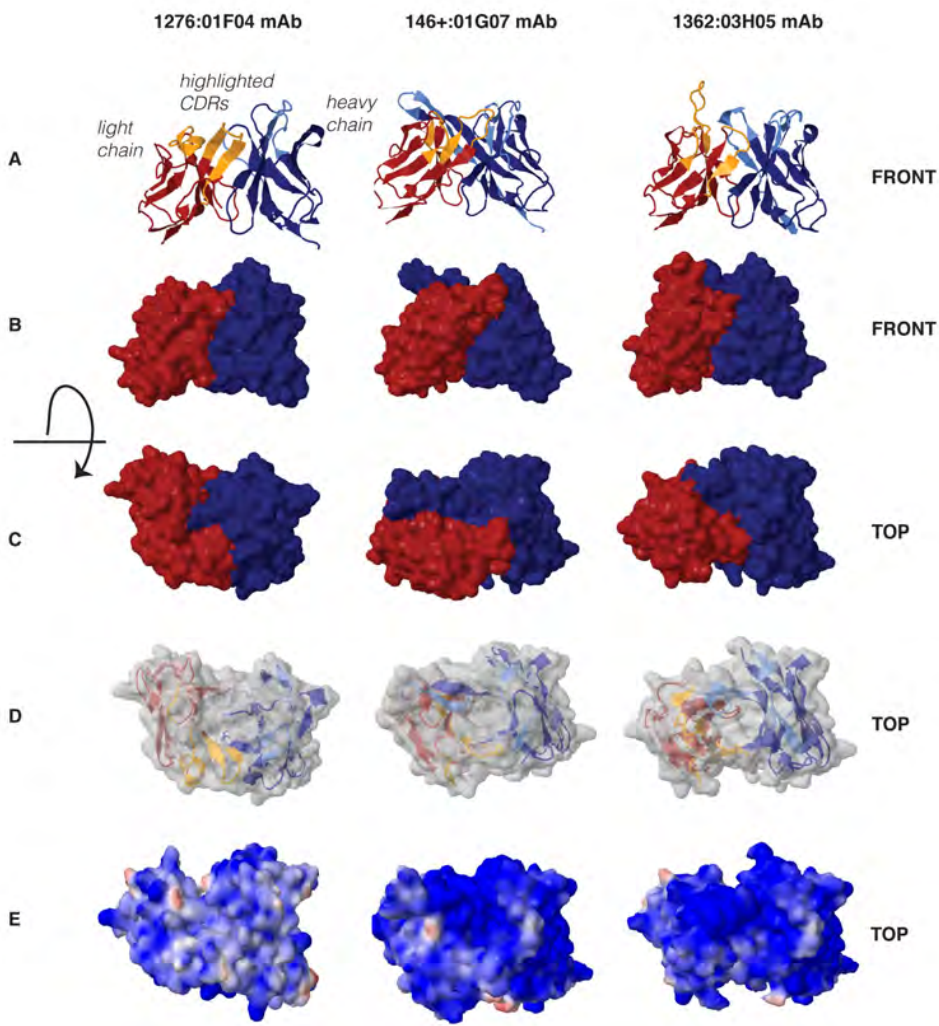

**Figure 7. Molecular modeling of the structure of variable region anti-MDA specific synovial antibodies**

Models are visualized for the VH/VL region of the mAbs 1276:01F04, 146+:01G07 and 1362:03H05. **A.** Cartoon model of light chain variable region in red and heavy chain region in blue with the light chain CDR loops are highlighted in orange and heavy chain CDR loops in light blue. **B-C.** Solid surface models showing the light chain in red and the heavy chain in blue from the side **(B)** and rotated 90 degrees to show the top **(C)**, looking into the potential antigen binding surface built up by the light and heavy chain CDR loops. **D.** Top view with translucent surface revealing a cartoon model of the secondary structure. **E.** Top view of molecular electrostatic potential (MEP) surfaces. The RWB color (scale -.5, .5) shows positively charged surfaces as blue and negatively charged as red. Structure models were generated with PIGS online tool using best H+L model and visualized with Jmol. Electrostatic calculations were calculated with the PDB2PQR server and Jmol MEP.

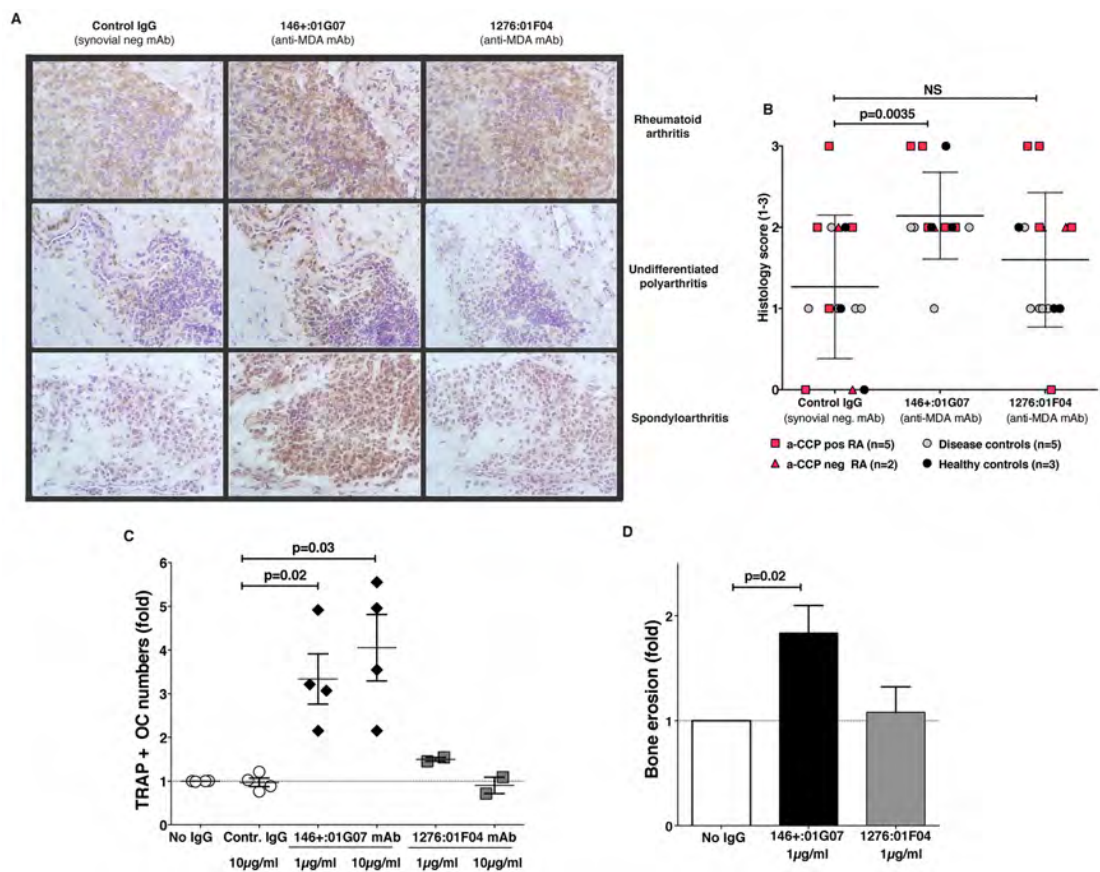

**Figure 8. Potential pathogenic properties of IgG anti-MDA is restricted to a subgroup of antibodies**

**A.** Representative results of immunohistochemistry staining of human synovial tissue with two anti-MDA mAbs 146+:01G07 and 1276:01F04, compared to a MDA/citrulline-negative plasma cell derived synovial monoclonal 1276:06D06 (Control IgG). Images are shown from x25 magnification. **B.** Summary of histology results from samples from seven RA patients (five seropositive patients and two seronegative patients), three healthy controls and five disease controls (including two spondyloarhtritis, one osteoarthritis with Sjögren's syndrome, and one psoriatic arthritis patients). The staining was scored by intensity 1-3 for the three monoclonal antibodies. Panels C-E show results of osteoclast stimulation. The monoclonal 146+:01G07 stimulates increased osteoclastogenesis in cultures of monocyte-derived human osteoclasts while the clone 1276:01F04 does not show any functional *in vitro* properties. Osteoclasts were generated *in vitro* from CD14-positive monocytes isolated from the circulation of healthy donors and stimulated to induce osteoclast differentiation with M-CSF and RANKL with or without the presence of human purified monoclonal IgG1 at indicated concentrations. **C.** Fold increase osteoclasts measured by counts of tartate-resistant acid phosphatase (TRAP) positive cells with ≥3 nuclei. **D.** Fold increase of osteoclasts by resorption area on calcium phosphate plates. The graphs show data from average of triplicates from two-four independent experiment using osteoclasts cultures from different donors. P-values from student's t-test are shown. No statistically significant difference was seen between 10 µg/ml and 1 µg/ml mAb IgG. The mAb 1362:01E02 was used as negative control.

**Table 1. Identification of malondialdehyde-modified proteins in RA synovial tissue by mass spectrometry**

| | | Number of identified MDA-peptides | | | | | | |
|---|---|---|---|---|---|---|---|---|
| | | Syn1 α-CCP+ RA | Syn2 α-CCP+ RA | Syn3 α-CCP+ RA | Syn4 α-CCP+ RA | Syn5 α-CCP+ RA | Syn6 α-CCP- RA | Syn7 α-CCP- RA |
| **MDA-proteins with Mascot score >13** | **Total No** | | | | | | | |
| Actin cytoplasmic 1 | 1 | 0 | 0 | 0 | 1 | 0 | 0 | 0 |
| Alpha-1-acid glycoprotein 1 | 1 | 0 | 0 | 0 | 1 | 0 | 0 | 0 |
| Apolipoprotein A-I | 1 | 0 | 0 | 0 | 1 | 0 | 0 | 0 |
| Hemoglobin subunit alpha | 6 | 0 | 0 | 4 | 6 | 0 | 1 | 0 |
| Hemoglobin subunit beta | 3 | 0 | 0 | 2 | 3 | 0 | 1 | 0 |
| Ig gamma-1 chain C region | 1 | 0 | 1 | 1 | 1 | 0 | 1 | 1 |
| Serotransferrin | 1 | 0 | 0 | 0 | 1 | 0 | 0 | 0 |
| Serum albumin | 11 | 0 | 4 | 4 | 8 | 0 | 2 | 2 |
| **Selected MDA-proteins with Mascot score <13** | | | | | | | | |
| Carbonic anhydrase 1 | 1 | 0 | 0 | 1 | 1 | 0 | 1 | 0 |
| Vimentin | 3 | 0 | 0 | 0 | 3 | 0 | 2 | 0 |
| **Total No:** | | 0 | 5 | 12 | 26 | 0 | 8 | 3 |

MDA-modification could be identified on lysine, histidine, arginine, glutamine or asparagine, for a complete list of identified MDA-peptides and Mascot scores see Supplemental Table 1.

At least one spectrum per peptide was manually confirmed to be accurate.

**Table 2. Anti-MDA and anti-PC natural antibody levels in different patient populations**

| | Controls (n=71) | OA (n=30) | | PsA (n=15) | | RA (n=162) | |
|---|---|---|---|---|---|---|---|
| | Mean±SD | Mean±SD | p-value# | Mean±SD | p-value# | Mean±SD | p-value# |
| IgM anti-MDA | 13±14 | 14±12 | 0.39 | 25±20 | **0.006** | 40±50 | **<0.0001** |
| IgM anti-PC | 26±28 | 39±33 | **0.007** | 40±40 | 0.12 | 55±44 | **<0.0001** |
| | | | | | | | |
| IgG anti-MDA | 5±3 | 7±6 | 0.06 | 7±4 | **0.01** | 13±11 | **<0.0001** |
| IgG anti-PC | 28±27 | 34±28 | 0.29 | 36±19 | 0.10 | 51±27 | **<0.0001** |

# P-values were derived from Mann-Whitney test compared to healthy controls
OA: osteoarthritis; PsA: psoriatic arthritis, RA: rheumatoid arthritis
MDA: Malondialdehyde-modification; PC: Phosphorylcholine

**Table 3. Antibody levels and disease activity in new onset RA compared to chronic RA**

| | NORA 0-6m (n=34) | ERA 6m-2y (n=7) | CRA >2y (n=56) | p-value NORA vs CRA[#] |
|---|---|---|---|---|
| IgM anti-MDA (RU/ ml, Mean ± SD) | 68±74 | 33±43 | 35±40 | **0.008** |
| IgM anti-PC (RU/ ml, Mean ± SD) | 62±49 | 60±34 | 52±41 | 0.44 |
| IgG anti-MDA (RU/ ml, Mean ± SD) | 21±13 | 8±5 | 10±8 | **<0.0001** |
| IgG anti-PC (RU/ ml, Mean ± SD) | 49±25 | 46±19 | 43±23 | 0.27 |
| IgG anti-CCP3 (RU/ ml, Mean ± SD) | 359±168 | 324±323 | 355±243 | 0.80 |
| DAS28-ESR (Mean ± SD) | 5.7±1.2 | 4.5±0.7 | 4.8±1.5 | **0.01** |

NORA: Disease onset 0-6 months
ERA: Disease onset 6 months-2 years
CRA: Disease onset >2years
# P-values were derived from Mann-Whitney test

**Table 4. Characteristics of the DMARD naïve RA cohort**

| Characteristics | |
|---|---|
| Age (years±SD) | 46±14 |
| Females (n, %) | 45, 73% |
| Ethnicity | |
|     White (n, %) | 44, 71% |
|     African Americans (n, %) | 8, 13% |
|     Asian (n, %) | 10, 16% |
| Hispanic (n, %) | 36, 58% |
| ESR (mm; mean±SD) | 40±29 |
| Patient global (mean±SD) | 55±27 |
| DAS28* (mean±SD) | 5.6±1.5 |
| CCP3 IgG pos. (n, %) | 54, 87% |
| IgM RF pos. (n, %) | 53, 85% |
| Seronegatives (n, %) | 7, 11% |
| IgG CCP3 (RU/ml; mean±SD) | 355±209 |
| IgM anti-MDA (RU/ml; mean±SD) | 30±36 |
| IgM anti-PC (RU/ml; mean±SD) | 56±49 |
| IgG anti-MDA (RU/ml; mean±SD) | 13±12 |
| IgG anti-PC (RU/ml; mean±SD) | 60±30 |

Characteristics for 62 DMARD naïve RA patients
* DAS28 calculations includes ESR

**Table 5. Association of IgG anti-MDA with inflammatory markers in DMARD naïve RA patients**

|  | Spearman R-value | p-value# |
|---|---|---|
| DAS28-ESR | 0.29 | **0.03** |
| Patient global | 0.096 | 0.51 |
| ESR | 0.39 | **0.002** |
| CRP | 0.37 | **0.003** |
| IL-17F | 0.26 | **0.05** |
| IL-1$\beta$ | 0.19 | 0.13 |
| IL-6 | 0.27 | **0.03** |
| TNF-$\alpha$ | 0.39 | **0.002** |
| VEGF | 0.06 | 0.68 |
| IL-6R$\alpha$ | -0.16 | 0.20 |
| sTNFRII | 0.13 | 0.28 |

Correlation analysis for 62 DMARD naïve RA patients
# P-values were derived from Spearman analysis

**Table 6. Antibody levels in RA patients with high disease activity compared to low disease activity**

| | Low/Remission *DAS28<3.2* *(n=14)** | | Moderate *DAS28 3.2-5.1* *(n=56)* | | High *DAS28 >5.1* *(n=87)* | |
|---|---|---|---|---|---|---|
| DAS28 (mean±SD) | 2.6±0.2 | | 4.3±0.5 | **p-value [#]** **<0.0001** | 6.4±0.8 | **p-value [#]** **<0.0001** |
| IgM anti-MDA (RU/ml; mean±SD) | 22±18 | | 32±34 | 0.23 | 48±60 | 0.15 |
| IgM anti-PC (RU/ml; mean±SD) | 47±33 | | 47±31 | 0.98 | 61±52 | 0.51 |
| IgG anti-MDA (RU/ml; mean±SD) | 6±3 | | 12±11 | **0.005** | 15±12 | **0.001** |
| IgG anti-PC (RU/ml; mean±SD) | 52±27 | | 51±25 | 0.95 | 51±29 | 0.88 |
| IgG anti-CCP3 (RU/ml; mean±SD) | 344±240 | | 344±203 | 0.96 | 364±209 | 0.87 |

*Seven patients were in remission DAS28<2.6 and seven patients had low disease activity DAS28 2.6-3.2
# P-values were derived by Mann-Whitney analysis compared to RA patients with low disease activity/DAS remission

**Table 7. Characteristics of synovial-derived monoclonal anti-MDA antibodies**

| CLONE-ID | B CELL TYPE | α-MDA | UNSPEC. | VL | JL | GERMLINE VL (%) | VL MUT. | VH | DH | JH | GERMLINE VH (%) | VH MUT. | HCDR3 |
|---|---|---|---|---|---|---|---|---|---|---|---|---|---|
| 1276:01F04 | Mem. IgG+ | ++++ | - | LV1-51*01 | LJ3*02 | 99.3 | 2 | 4-39*01 | 3-10*02 | 4*02 | 99.7 | 1 | VRVRGYFDY |
| 146+:01G07 | Mem. FcRL4+ IgA+ | ++ | - | KVD3-15*01 | KJ2*01 | 93.5 | 18 | 3-33*01 | 5-24*01 | 4*02 | 99.7 | 1 | ARARRGDGYNQARYYYFDY |
| 1103:01H05 | Mem. IgG+ | ++ | + | LV1-40*01 | LJ2*01 | 96.6 | 10 | 3-30*01 | 1-26*01 | 4*02 | 100 | 0 | ARDPYRGKATQDY |
| 1103:01A03 | Mem. IgG+ | ++ | + | LV1-51*01 | LJ3*02 | 98.0 | 6 | 4-39*01 | 5-5*01 | 4*02 | 97.3 | 8 | ARRRGYSYGYSRARGTTFDY |
| 1362:03H05 | Plasma IgG+ | ++ | - | KV4-1*1 | KJ2*01 | 95.9 | 12 | 3-53*01 | 3-3*02 | 6*02 | 95.5 | 13 | ARDRRGWSGYYSLRYGMDV |
| 1362:07D01 | Plasma IgG+ | ++ | +++ | KV1-39 | KJ2*01 | 95.8 | 12 | 4-34*1 | 2-8*01 | 6*02 | 95.9 | 12 | AREGTWPPYNYYYFGMDV |

FcRL4: Fc Receptor like 4

**Supplemental Table 1. Identification of MDA-modified proteins in RA synovial tissue by mass spectrometry**

| MDA modified peptides | | | | MASCOT SCORES | | | | | | | |
|---|---|---|---|---|---|---|---|---|---|---|---|
| **Accession** | **Description** | **Peptide sequence** | **MDA-site** | **Max score** | **Syn1** | **Syn2** | **Syn3** | **Syn4** | **Syn5** | **Syn6** | **Syn7** |
| ACTB_HUMAN[1] | Actin cytoplasmic 1 | RGILTLK | (R1) | 16.24 | | | | 16.24 | | | |
| A1AG1_HUMAN | α1-acid glycoprotein 1 | NWGLSVYADKPETTK | (N1) | 47.27 | | | | 47.27 | | | |
| APOA1_HUMAN | Apolipoprotein A-I | DSGRDYVSQFEGSALGK | (R4) | 17.35 | | | | 17.35 | | | |
| CAH1_HUMAN[2] | Carbonic anhydrase 1 | HDTSLKPISVSYNPATAK | (H1) | *10.03* | | | 10.03 | 5.77 | | 6.19 | |
| HBA_HUMAN[3] | Hemoglobin subunit α | KVADALTNAVAHVDDMPNALSALSD | (K1) | 21.45 | | | 21.45 | 15.2 | | | |
| HBA_HUMAN[3] | | LLSHCLLVTLAAHLPAEFTPAVHASL | (H4) | 24.77 | | | | 24.77 | | | |
| HBA_HUMAN[3] | | LRVDPVNFK | (R2) | 14.23 | | | | 14.23 | | | |
| HBA_HUMAN[3] | | TYFPHFDLSHGSAQVK | (H5) | 15.42 | | | 15.42 | 14.43 | | | |
| HBA_HUMAN | | VGAHAGEYGAEALERMFLSFPTTK | (R15) | 25.08 | | | 25.08 | 13.07 | | | |
| HBA_HUMAN | | VGAHAGEYGAEALERMFLSFPTTK | (H4) | 62.01 | | | 36.65 | 62.01 | | 49.32 | |
| HBB_HUMAN[4] | Hemoglobin subunit β | KVLGAFSDGLAHLDNLK | (K1) | 61.51 | | | 58.90 | 61.51 | | 57.49 | |
| HBB_HUMAN[5] | | LHVDPENFRLLGNVLVCVLAHHFGK | (H2) | 24.79 | | | 24.02 | 24.79 | | | |
| HBB_HUMAN[5] | | VVAGVANALAHKYH | (N7) | 14.49 | | | | 14.49 | | | |
| IGHG1_HUMAN | Ig γ-1 chain C region | FNWYVDGVEVHNAK | (N2) | 33.37 | | 19.88 | 23.61 | 33.37 | | 30.51 | 13.75 |
| TRFE_HUMAN[6] | Serotransferrin | HQTVPQNTGGK | (H1) or (Q2) | 16.09 | | | | 16.09 | | | |
| ALBU_HUMAN[7] | Serum albumin | AWAVARLSQRFPKAEFAEVSK | (K13) | 25.56 | | | | 25.56 | | 19.26 | |
| ALBU_HUMAN[8] | | FQNALLVRYTK | (R8) | 32.7 | | | 27.09 | 26.25 | | | 32.7 |
| ALBU_HUMAN[8] | | FQNALLVRYTK | (N3) | 24.62 | | 24.62 | | | | | |
| ALBU_HUMAN[8] | | FQNALLVRYTK | (Q2) | 23.8 | | | | | | | 23.8 |
| ALBU_HUMAN[8] | | KVPQVSTPTLVEVSRNLGK | (K1) | 30.00 | | 18.95 | 17.35 | 30 | | | |
| ALBU_HUMAN[8] | | KVPQVSTPTLVEVSRNLGK | Q4) | 15.41 | | 15.41 | | | | | |
| ALBU_HUMAN[8] | | QNCELFEQLGEYK | (N2) | 31.79 | | | 31.79 | 31.05 | | 20.32 | |
| ALBU_HUMAN[8] | | QTALVELVK | (Q1) | 18.72 | | | 18.72 | | | | |
| ALBU_HUMAN[8] | | RMPCAEDYLSVVLNQLCVLHEK | (R1) | 48.51 | | | | 48.51 | | | |
| ALBU_HUMAN[8] | | VFDEFKPLVEEPQNLIK | (K6) | 13.06 | | | | 13.06 | | | |
| ALBU_HUMAN[8] | | VHTECCHGDLLECADDRADLAK | (H2) | 28.22 | | 17.63 | | 28.22 | | | |
| VIME_HUMAN | Vimentin | RTLLIK | (R1) | *1.83* | | | | 0.71 | | 1.83 | |
| VIME_HUMAN | | TVETRDGQVINETSQHHDDLE | (R5) | *5.22* | | | | 5.22 | | | |
| VIME_HUMAN[9] | | VRFLEQQNK | (R2) | *12.05* | | | | 2.18 | | 12.05 | |

At least one spectrum per peptide was manually confirmed to be accurate.
Peptides with maximum mascot score less than 13 are highlighted in grey.

***Alternative accession numbers:***

1. CTG_HUMAN, ACTA_HUMAN, ACTH_HUMAN, C9JZR7_HUMAN, ACTC_HUMAN, ACTS_HUMAN, Q5T8M8_HUMAN, A6NL76_HUMAN, POTEE_HUMAN, POTEF_HUMAN, POTEI_HUMAN, POTEJ_HUMAN, C9JFL5_HUMAN, Q5T8M7_HUMAN, I3L3I4_HUMAN
2. E5RH81_HUMAN, E5RHP7_HUMAN
3. G3V1N2_HUMAN
4. HBD_HUMAN, E9PFT6_HUMAN
5. HBD_HUMAN
6. J3KN47_HUMAN
7. B7WNR0_HUMAN
8. B7WNR0_HUMAN, C9JKR2_HUMAN
9. K22O_HUMAN, K2C8_HUMAN, F8VU64_HUMAN, F8VUG2_HUMAN, K2C7_HUMAN, F8VRG4_HUMAN, K2C6A_HUMAN, K2C6B_HUMAN, K2C6C_HUMAN, KRT85_HUMAN, K2C3_HUMAN, J3QST3_HUMAN, K2C75_HUMAN, K2C79_HUMAN, K2C5_HUMAN, KRT84_HUMAN, GFAP_HUMAN, B0YJC4_HUMAN, KRT83_HUMAN, KRT81_HUMAN, KRT86_HUMAN, K2C78_HUMAN, F8VZY5_HUMAN, B4DIR1_HUMAN

## Supplemental Table 2. Mass spectrometry analysis of *in vitro* modified BSA

| MDA modified peptides | | MASCOT SCORES* | | | |
|---|---|---|---|---|---|
| Sequence (bovine albumin ALBU_BOVIN) | MDA- Site | Max score | BSA (ctrl) | BSA (MDA 2h) | BSA (MDA 24h) |
| ADLAKYICDNQDTISSK | MDA (K5) | 21.16 | | 13.28 | 21.16 |
| AEFVEVTKLVTDLTK | MDA (K8) | 51.98 | | 51.98 | 29.25 |
| AFDEKLFTFHADICTLPDTEK | MDA (K5) | 17.44 | | | 17.44 |
| ALKAWSVAR | MDA (K3) | 17.72 | | 15.29 | 17.72 |
| ATEEQLKTVMENFVAFVDK | MDA (K7) | 20.2 | | 16.42 | 20.2 |
| CCAADDKEACFAVEGPK | MDA (K7) | 38.36 | | 38.36 | 18.67 |
| EACFAVEGPKLVVSTQTALA | MDA (K10) | 14.37 | | | 14.37 |
| FPKAEFVEVTK | MDA (K3) | 32.73 | | 32.73 | 20.36 |
| HLVDEPQNLIKQNCDQFEK | MDA (K11) | 23.8 | | 23.8 | |
| HPEYAVSVLLR | MDA (H1) | 13.21 | | | 13.21 |
| KVPQVSTPTLVEVSR | MDA (K1) | 34.59 | 17.4 | 23.81 | 34.59 |
| LAKEYEATLEECCAK | MDA (K3) | 43.53 | | 43.53 | 36.56 |
| LCVLHEKTPVSEK | MDA (K7) | 27.56 | | | 27.56 |
| LSQKFPKAEFVEVTK | MDA (K7) | 26.28 | | 26.28 | 20.67 |
| LVNELTEFAKTCVADESHAGCEK | MDA (K10) | 23.9 | | 23.9 | |
| NYQEAKDAFLGSFLYEYSR | MDA (K6) | 22.03 | | 20.34 | 22.03 |
| QNCDQFEKLGEYGFQNALIVR | MDA (K8) | 33.67 | | 13.15 | 33.67 |
| QNCDQFEKLGEYGFQNALIVR | MDA (K8), Gln->pyro-Glu (N-term Q) | 16.67 | | | 16.67 |
| RHPEYAVSVLLR | MDA (R1 or H2) | 17.22 | | 13.75 | 17.22 |
| RHPYFYAPELLYYANK | MDA (R1 or H2) | 15.72 | 15.72 | | |
| SLHTLFGDELCKVASLR | MDA (K12) | 23.32 | | 21.06 | 23.32 |
| TCVADESHAGCEKSLHTLFGDELCK | MDA (K13) | 16 | | | 16 |
| TPVSEKVTKCCTESLVNR | MDA (K6), MDA (K9) | 15.49 | | 15.49 | 13.95 |
| TPVSEKVTKCCTESLVNR | MDA (K9) | 14.36 | | | 14.36 |
| TVMENFVAFVDKCCAADDK | MDA (K12) | 17.1 | | | 17.1 |
| TVMENFVAFVDKCCAADDKEACFAVEGPK | MDA (K19) | 30.04 | | 30.04 | 21.91 |
| TVMENFVAFVDKCCAADDKEACFAVEGPK | MDA (K12) | 17.48 | | 17.48 | |
| TVMENFVAFVDKCCAADDKEACFAVEGPK | MDA (K12), MDA (K19) | 23.2 | | 23.2 | 14.76 |
| TVMENFVAFVDKCCAADDKEACFAVEGPK | Oxidation (M3), MDA (K12) | 17.52 | | 17.52 | |
| TVMENFVAFVDKCCAADDKEACFAVEGPK | Deamidated (N5), MDA (K12) | 14.36 | | 14.36 | |
| TVMENFVAFVDKCCAADDKEACFAVEGPK | Deamidated (N5), MDA (K19) | 26.11 | | | 26.11 |
| VTKCCTESLVNR | MDA (K3) | 36.91 | | 36.91 | 31.64 |
| YICDNQDTISSKLK | MDA (K12) | 19.29 | | 19.29 | 13.42 |

| MAA modified peptides | | MASCOT SCORES* | | | |
|---|---|---|---|---|---|
| Sequence | MAA-Site | Max score | BSA (ctrl) | BSA (MDA 2h) | BSA (MDA 24h) |
| ADLAKYICDNQDTISSK | MAA (K5) | 27.47 | | | 27.47 |
| ALKAWSVAR | MAA (K3) | 20.24 | | | 20.24 |
| CASIQKFGER | MAA (K6) | 18.13 | | | 18.13 |
| CCTKPESERMPCTEDYLSLILNR | MAA (K4) | 15.01 | | | 15.01 |
| DTHKSEIAHR | MAA (K4) | 13.65 | | | 13.65 |
| ETYGDMADCCEKQEPER | MAA (K12) | 43.28 | | | 43.28 |
| KVPQVSTPTLVEVSR | MAA (K1) | 29.81 | | 17.36 | 29.81 |
| LAKEYEATLEECCAK | MAA (K3) | 25.74 | | | 25.74 |
| LVNELTEFAKTCVADESHAGCEK | MAA (K10) | 19.66 | | | 19.66 |
| NYQEAKDAFLGSFLYEYSR | MAA (K6) | 44.6 | | | 44.6 |
| SLHTLFGDELCKVASLR | MAA (K12) | 24.27 | | | 24.27 |
| TVMENFVAFVDKCCAADDKEACFAVEGPK | MAA (K12) | 25.35 | | | 25.35 |
| TVMENFVAFVDKCCAADDKEACFAVEGPK | Deamidated (N5), MAA (K12) | 19.54 | | | 19.54 |
| VTKCCTESLVNR | MAA (K3) | 32.88 | | | 32.88 |

*limit for significance = 13
Each column shows the highest score from 4 technical replicate MS analysis.
At least one spectrum per peptide was manually confirmed to be accurate.

**Supplemental Table 3. Characteristics of CRA with standard care compare to DMARD naïve CRA**

| Characteristics | CRA | DMARD naïve CRA | p-value[#] |
|---|---|---|---|
| Age (years±SD) | 54±10 | 50±14 | 0.44 |
| Females (n, %) | 23, 82% | 24, 85% | 1.0 |
| RA duration | 15±12 | 9±8 | **0.004** |
| DAS28 | 5.2±1.6 | 5.3±1.6 | 0.79 |
| CCP3 IgG pos. (n, %) | 26, 93% | 20, 71% | 0.08 |
| IgM RF (n, %) | 26, 93% | 20, 71% | 0.08 |
| Seronegative (n, %) | 2, 7% | 8, 29% | 0.08 |
| IgG CCP3 (RU/ml; mean±SD) | 367±237 | 277±220 | 0.09 |
| IgM anti-MDA (RU/ml; mean±SD) | 35±34 | 22±24 | 0.09 |
| IgM anti-PC (RU/ml; mean±SD) | 54±40 | 46±42 | 0.20 |
| IgG anti-MDA (RU/ml; mean±SD) | 10±8 | 13±13 | 0.55 |
| IgG anti-PC (RU/ml; mean±SD) | 42±24 | 55±31 | 0.12 |

N=28 in both groups, disease duration > 2 years
#P-values were derived by Mann-Whitney analysis

**Supplemental Table 4. Screened antibody clones derived from single cell sorted synovial B cells**

| Patient | Patient status | Source of B cells | Synovial B cell subset | NO of screened clones | NO of anti-MDA positive clones |
|---------|----------------|-------------------|------------------------|-----------------------|--------------------------------|
| RA146 | CCP pos | Synovial tissue | FcRL4 + | 18 | 1 |
| RA146 | CCP pos | Synovial tissue | FcRL4 - | 11 | |
| RA153 | CCP pos | Synovial fluid | FcRL4 + | 10 | |
| RA153 | CCP pos | Synovial fluid | FcRL4 - | 10 | |
| RA1003 | CCP pos | Synovial fluid | IgG+ memory | 7 | |
| RA1103 | CCP pos | Synovial fluid | IgG+ memory | 17 | 2 |
| RA1124 | CCP pos | Synovial fluid | IgG+ memory | 2 | |
| RA1276 | CCP pos | Synovial fluid | IgG+ memory | 20 | 1 |
| RA1276 | CCP pos | Synovial fluid | ASC | 33 | |
| RA1444 | CCP pos | Synovial fluid | IgG+ memory | 7 | |
| RA1444 | CCP pos | Synovial fluid | ASC | 9 | |
| RA1325 | CCP pos | Synovial fluid | IgG+ memory | 7 | |
| RA1325 | CCP pos | Synovial fluid | ASC | 2 | |
| RA0663 | CCP neg | Synovial fluid | IgG+ memory | 2 | |
| RA1362 | CCP neg | Synovial fluid | IgG+ memory | 3 | |
| RA1362 | CCP neg | Synovial fluid | ASC | 42 | 2 |
| | | | Total number of screened clones: | 200 | |

ACS: Antibody secreting cell; FcRL4: Fc Receptor like 4

**Supplemental Table 5. Citrullinated peptides in the ISAC microarray**

| Peptide | Protein | Amino acids | Amino acid sequence# | Reference |
|---|---|---|---|---|
| Cit-Fil$_{307-324}$ | Filaggrin | 307-324 | SHQEST(cit)GRSRGRSGRSGS (cyclic) | (1, 2) |
| Cit-Vim$_{60-75}$ | Vimentin | 60-75 | VYAT(cit)SSAV(cit)L(cit)SSVP | (3) |
| Cit-Vim$_{2-17}$ | Vimentin | 2-17 | ST(cit)SVSSSSY(cit)(cit)MFGG | (4) |
| Cit-Fibβ$_{36-52}$ | Fibrinogen β-chain | 36-52 | NEEGFFSA(cit)GHRPLDKK | (3) |
| Cit-Fibα$_{563-583}$ | Fibrinogen α-chain | 563-583 | HHPGIAEFPS(cit)GKSSSYSKQF | (5) |
| Cit-Fibα$_{580-600}$ | Fibrinogen α-chain | 580-600 | SKQFTSSTSYN(cit)GDSTFESKS | (5) |
| Cit-Fibβ$_{(Cit-R)62-81a}$ | Fibrinogen β-chain | 62-81 | APPPISGGGY(cit)ARPAKAAAT | (5) |
| Cit-Fibβ$_{(R-Cit)62-81b}$ | Fibrinogen β-chain | 62-81 | APPPISGGGYRA(cit)PAKAAAT | (5) |
| CEP-1$_{5-21}$ | α-Enolase | 5-21 | CKIHA(cit)EIFDS(cit)GNPTVEC (cyclic) | (6) |

#Citrullinated (cit) peptides used in the Phadia's ImmunoCAP ISAC microarray system (7)

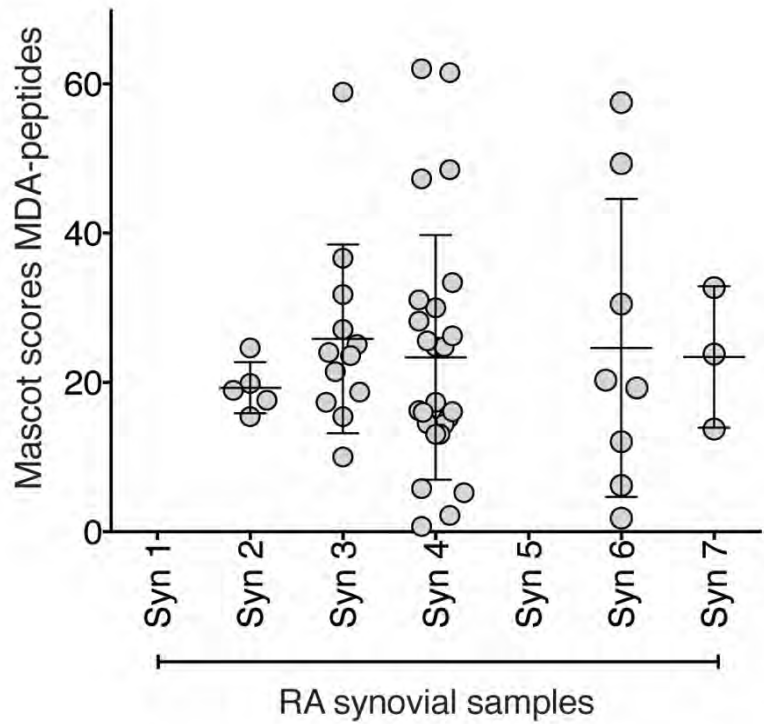

**Supplemenatal Figure 1. Identification of malondialdehyde-modified proteins in RA synovial tissue by mass spectrometry**

The figure depicts Mascot scores for MDA-containing peptides identified by mass spectrometry in RA synovial tissue. MDA-modification could be identified on lysine, histidine, arginine, glutamine or asparagine, for a complete list of identified MDA-peptides and Mascot scores see Supplemental Table 1. At least one spectrum per peptide was manually confirmed to be accurate. Patient 1-5 were seropositive for ACPA and patient 6-7 were seronegative.

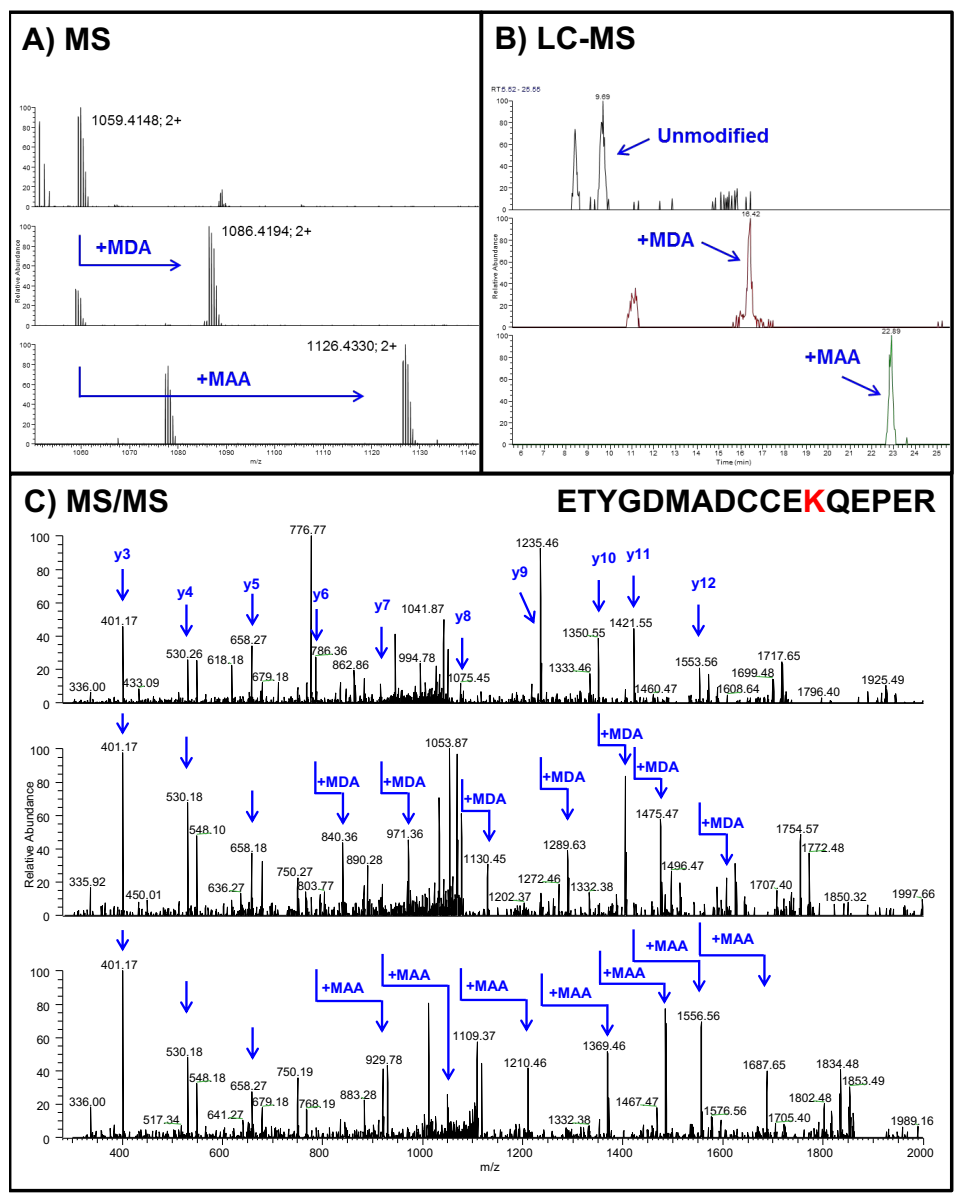

**Supplemental Figure 2. Example of mass spectrometry spectra for identification of malondialdehyde (MDA) and malondiadehyde/acetaldehyde (MAA) modified peptides.**

The figure shows a tryptic peptide from 24h MDA-modified bovine serum albumin (BSA), ETYGDMADCCEKQEPER. Three different forms of the peptide was identified: unmodified (upper trace in each spectrum), MDA modified (middle trace) and MAA (lower trace). **A**) shows the mass spectrum for the three forms indicating the mass shifts of 54.0106 Da for MDA and 134.0368 for MAA. **B**) shows the shifts in retention time for the three different forms. **C**) show the mass shifts in the fragments that contain the modifications. For clarity, only y3 to y12 have been indicated. The modified site (K) has been marked in red. All identified modified peptides in the MDA-modified BSA are presented in Supplemental Table 2.

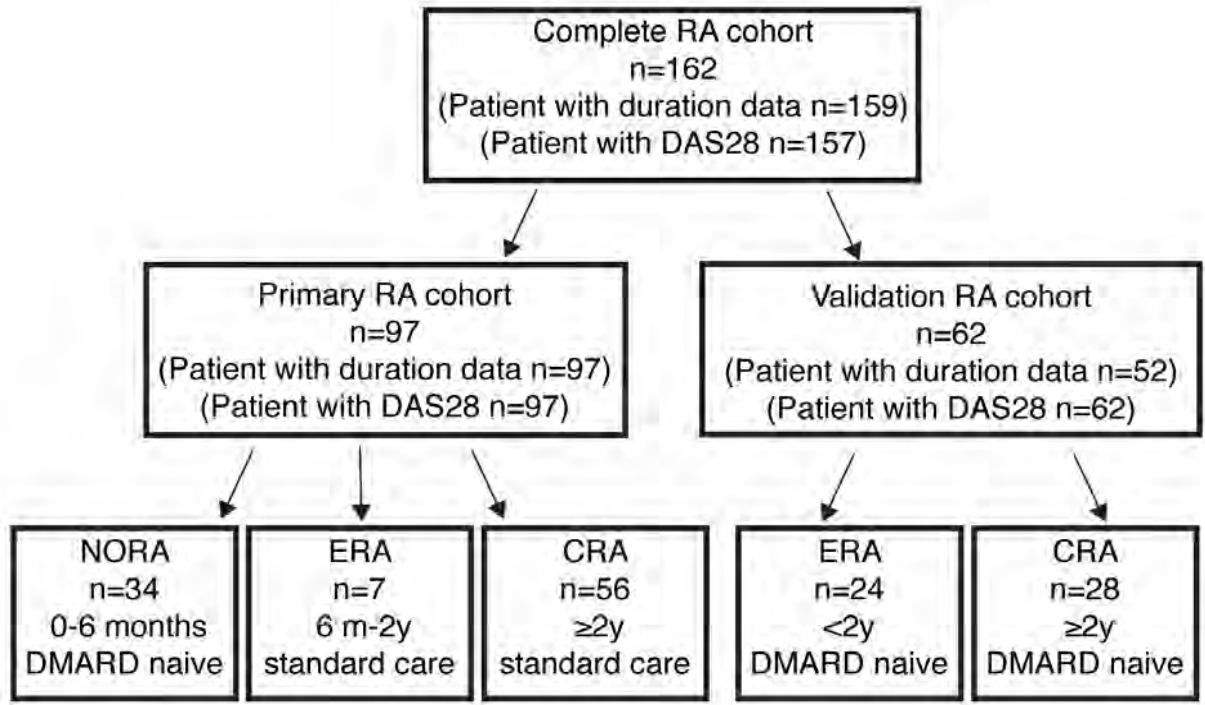

**Supplemental Figure 3. Schematic overview of RA patient samples included in serological anti-MDA studies**

Serological studies and statistical analysis were performed either on the primary RA cohort, the secondary DMARD naïve cohort, or the complete cross-sectional RA patient cohort.

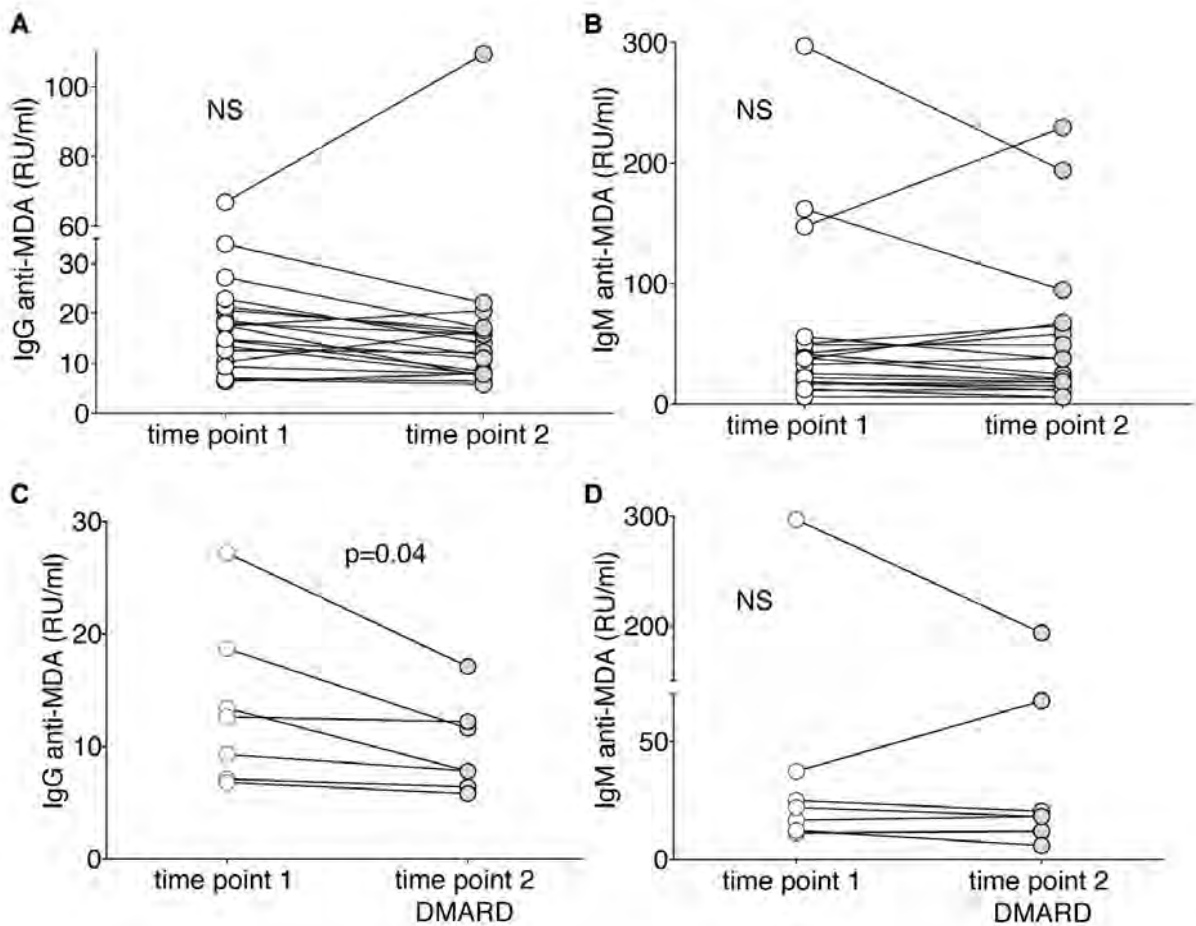

**Supplemental Figure 4. Variation of IgG anti-MDA antibody levels over time.**
ELISA screeing of IgG (**A, C**) and IgM (**B, D**) anti-MDA serum levels in 19 new onset RA patients (NORA) at two differen time points with 2-4 weeks in between. In most patients the levels remained relativly constant and there were no significant change. However, 7 of these patients did receive DMARD treatment (methotrexate, steroids, or plaquenil) between the two time points, presented in panel **C-D**, and for these patients the level of IgG anti-MDA did weakly but significantly decreased. Evaluated with paired t-test.

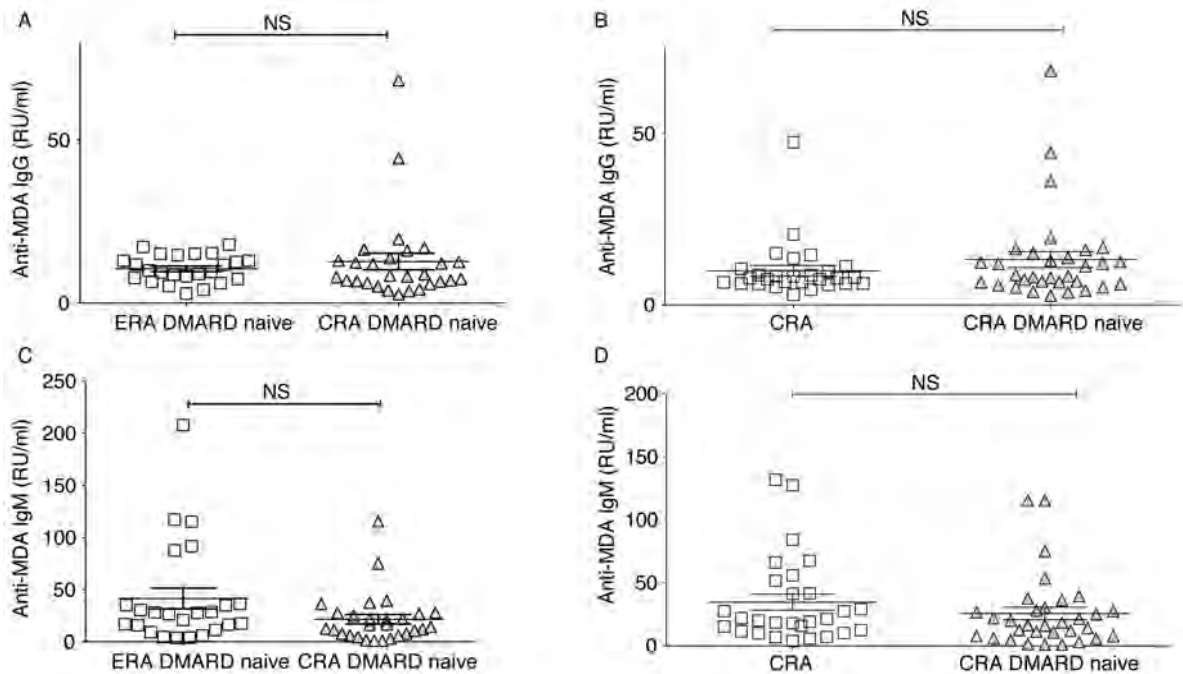

**Supplemental Figure 5. Anti-MDA autoantibody levels were not significantly different in chronic patients with similar disease activity but with or without DMARD treatment.**

No difference was seen in anti-MDA levels between patient groups in the DMARD naïve cohort. **A, C**. Anti-MDA IgG and IgM in DMARD naïve early RA (ERA, disease duration <2yrs, n=24) and chronic RA (CRA, disease duration ≥2yrs, n=28). **D, E**. Anti-MDA antibodies in chronic RA receiving standard care (n=28, DAS28 5.2±1.6) and DMARD naïve chronic RA (n=28, DAS28 5.3±1.6).

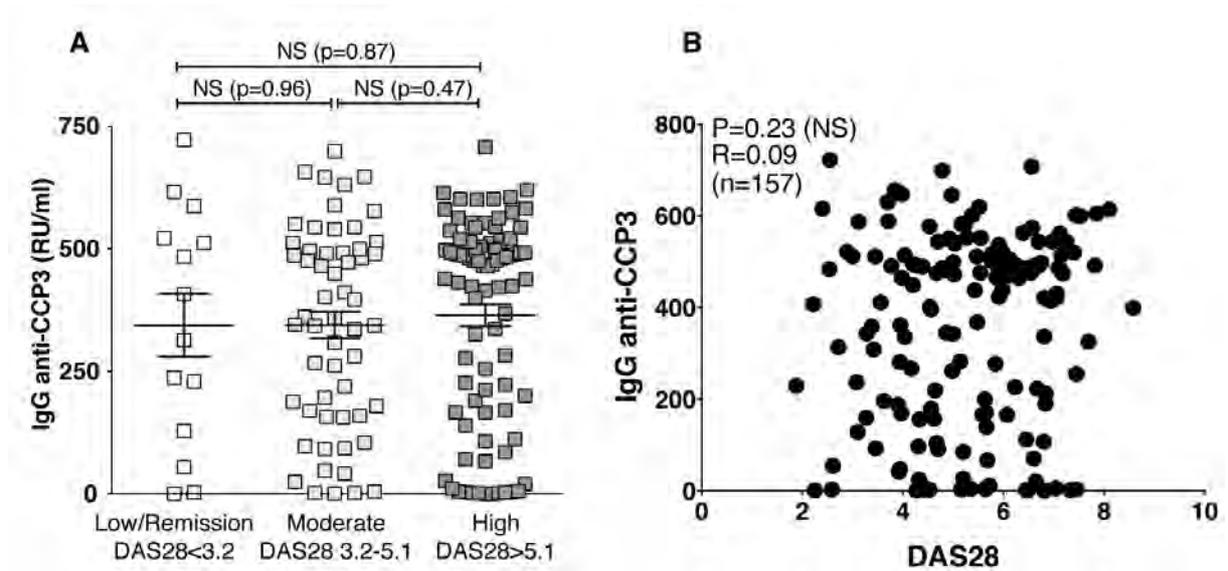

**Supplemental Figure 6. Serum anti-CCP3 levels are not associated with disease activity by DAS28**
**A**. Serum anti-CCP3 by ELISA in RA patients with low disease activity (DAS28 ESR< 3.2, n=14), moderate disease activity (DAS38 ESR 3.2-5.1, n=56) or high disease activity (DAS28>5.1, n=87). P-values were derived from Mann-Whitney analysis. **B**. Spearman association between DAS28 ESR scores and IgG anti-CCP3 levels.

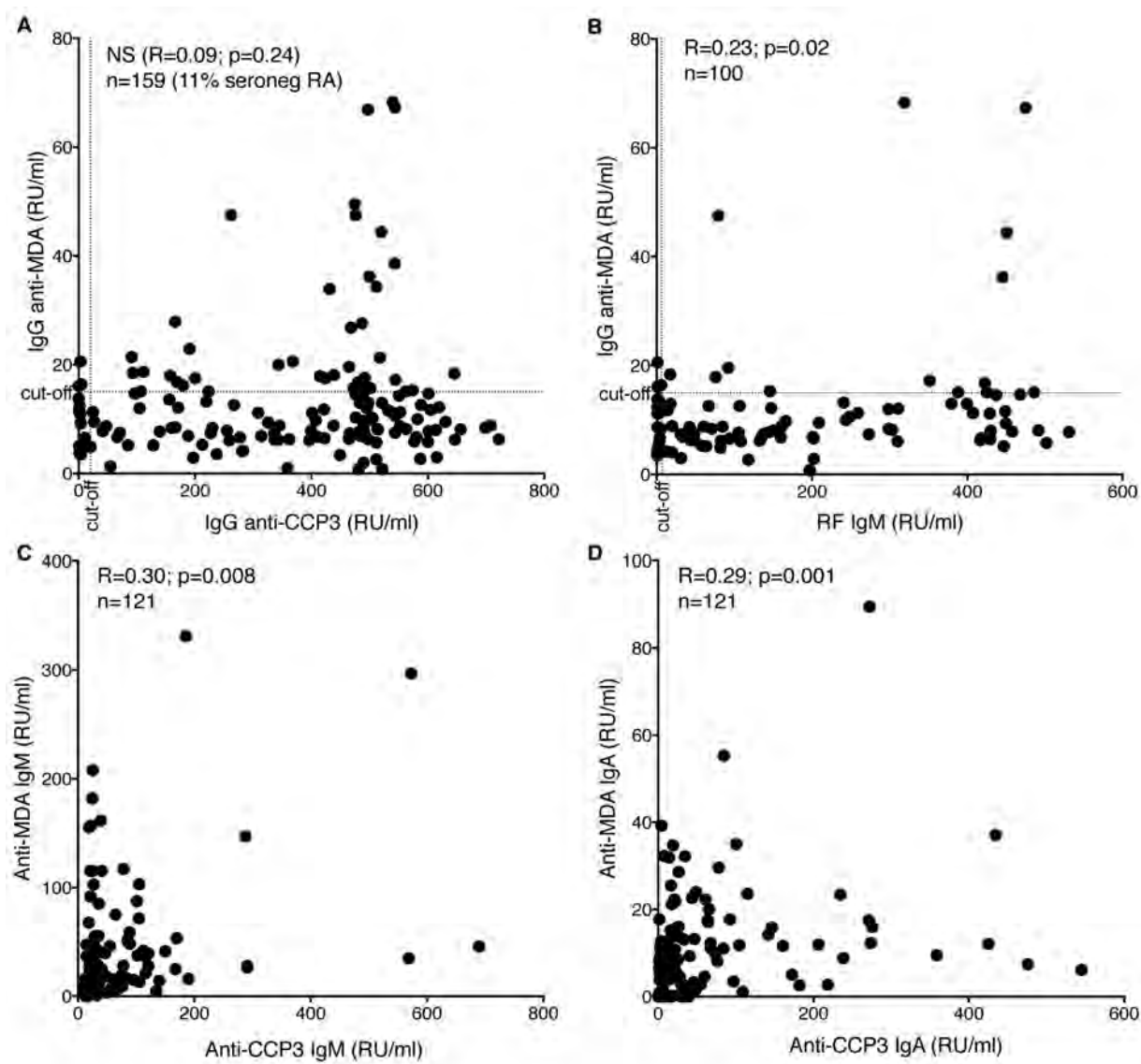

**Supplemental Figure 7. Association of serum levels of IgG anti-MDA with anti-CCP3 or RF**

IgG anti-CCP3 and IgM RF levels were determined by commercial ELISA kits (Inova Diagnostics) while anti-MDA IgG, IgM, and IgA as well as CCP3 IgA and IgM were measured with in-house ELISAs. Cut-off for positivity when available is marked with a dotted line. Cut-off for IgG anti-MDA (15 RU/ml) was determined by values for healthy controls (n= Mean + 3SD) while the cut-off for IgG anti-CCP3 (20RU/ml) and RF (6RU/ml) was based on the recommendation from the manufacturer of the clinical tests used (Inova Diagnostics). **A.** IgG anti-MDA vs IgG anti-CCP3 **B.** IgG anti-MDA vs IgM RF. **C.** IgM anti-MDA vs IgM anti-CCP3 **D.** IgA anti-MDA vs IgA anti-CCP3. IgG anti-MDA did not correlate with IgG anti-CCP3. There was a weak significantly correlation between IgG anti-MDA levels and RF IgM levels. Similarly, IgA anti-MDA correlated with IgA anti-CCP3 and IgM anti-MDA correlated with IgM anti-CCP3. P-values were derived from Spearman correlation analysis.

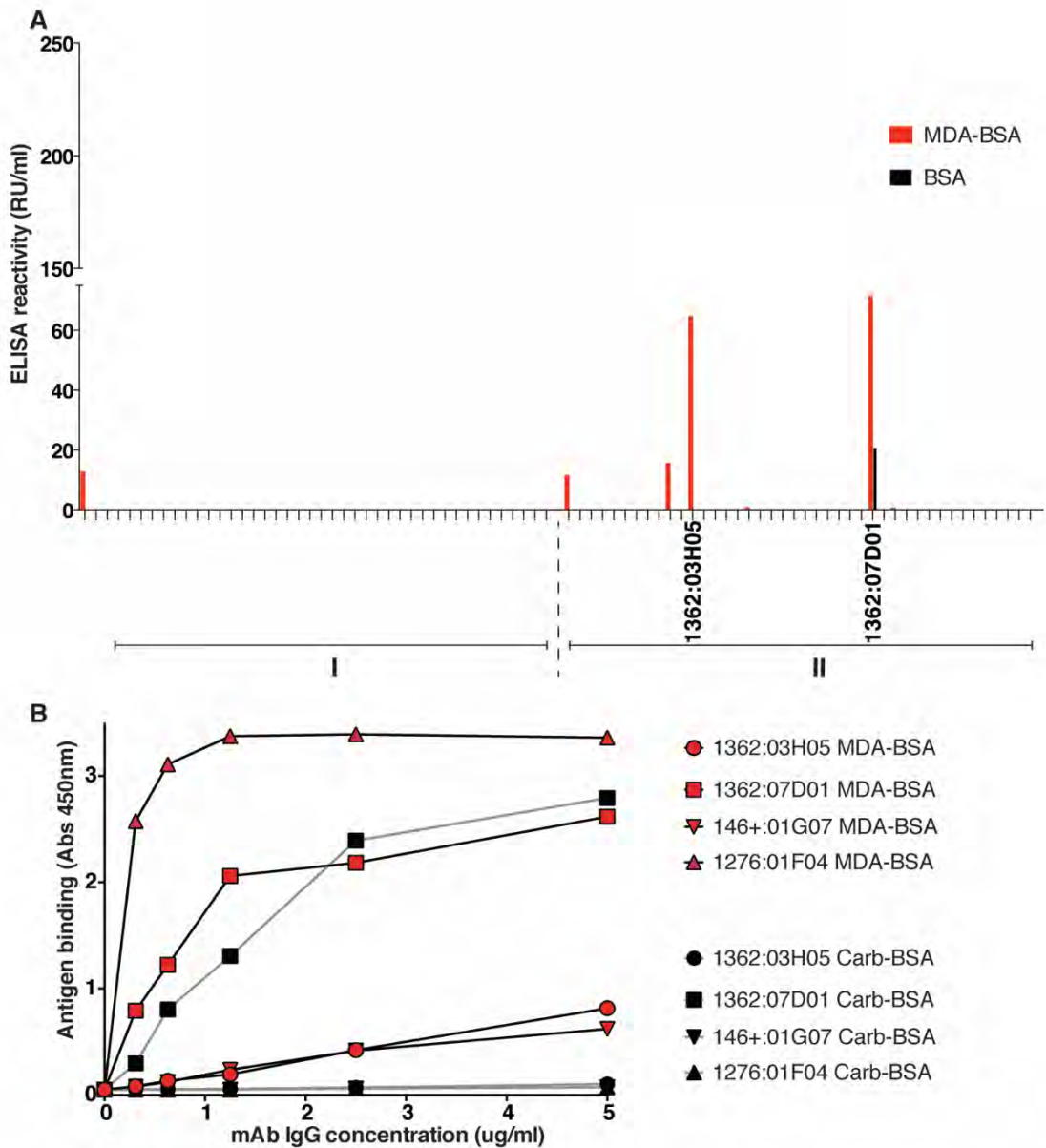

**Supplemental Figure 8. Identification of MDA-reactive antibody clones derived from synovial plasma cells**

ELISA screening of purified recombinant human monoclonal antibodies isolated from synovial antibody secreting cells (ASC) identified by the fluorescent foci method, cloned, and expressed in Expi293 cells as IgG1. **A.** 84 ASC mAbs were screened at 5 ug/ml for binding to MDA-BSA and to control wells coated with unmodified BSA. Reactivity was quantified as Relative Units (RU)/ml based on a standard reference curve. Antibodies were derived from different synovial subsets: *I.* IgG secreting cells from CCP positive patients (n=42) *II*. IgG secreting cells from CCP negative patients (n=42). **B.** Two mAbs showed significant MDA reactivity (1362:03H05 and 1362:07D01) and further evaluated for MDA-BSA binding in serial dilution compared to reactivity to the control antigen carbamylated BSA (Carb-BSA). In the graph the ASC mAbs are compared to the memory B cell identified clone 146+:01G07 and 1376:01F04. The clone 1362:07D01 shows high level of polyreactivity/unspecific binding while the clone 1362:03H05 is MDA-specific.

**A**

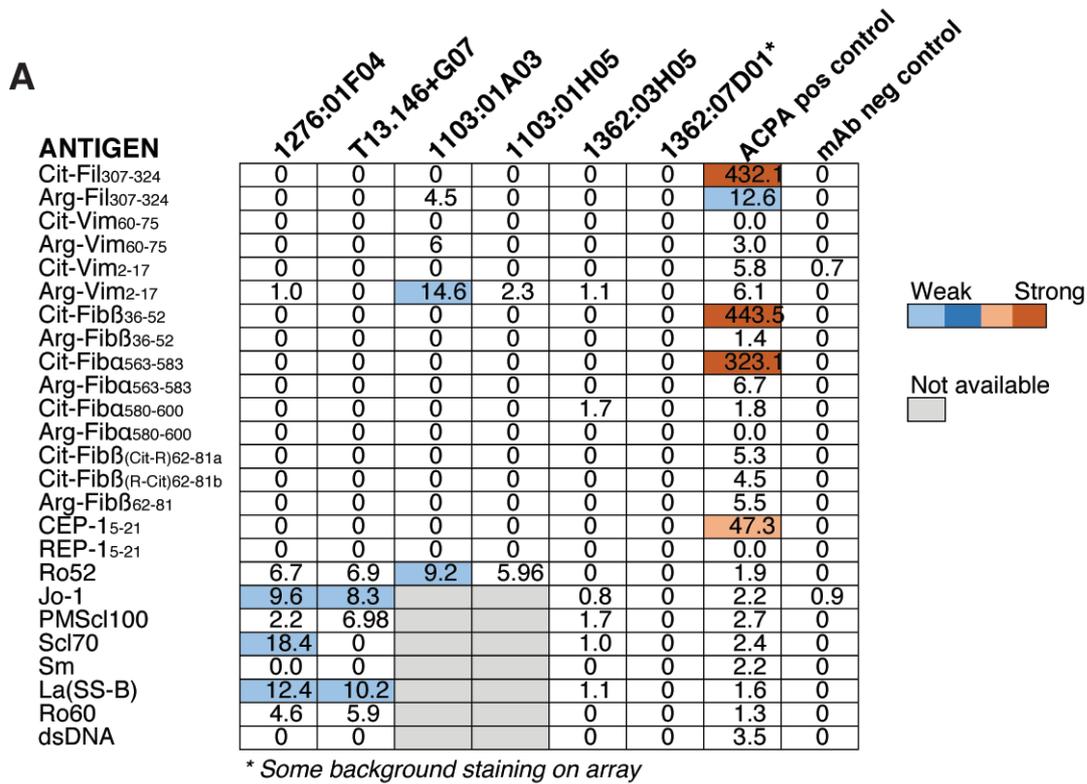

| ANTIGEN | 1276:01F04 | T13.146+G07 | 1103:01A03 | 1103:01H05 | 1362:03H05 | 1362:07D01* | ACPA pos control | mAb neg control |
|---|---|---|---|---|---|---|---|---|
| Cit-Fil$_{307-324}$ | 0 | 0 | 0 | 0 | 0 | 0 | 432.1 | 0 |
| Arg-Fil$_{307-324}$ | 0 | 0 | 4.5 | 0 | 0 | 0 | 12.6 | 0 |
| Cit-Vim$_{60-75}$ | 0 | 0 | 0 | 0 | 0 | 0 | 0.0 | 0 |
| Arg-Vim$_{60-75}$ | 0 | 0 | 6 | 0 | 0 | 0 | 3.0 | 0 |
| Cit-Vim$_{2-17}$ | 0 | 0 | 0 | 0 | 0 | 0 | 5.8 | 0.7 |
| Arg-Vim$_{2-17}$ | 1.0 | 0 | 14.6 | 2.3 | 1.1 | 0 | 6.1 | 0 |
| Cit-Fibß$_{36-52}$ | 0 | 0 | 0 | 0 | 0 | 0 | 443.5 | 0 |
| Arg-Fibß$_{36-52}$ | 0 | 0 | 0 | 0 | 0 | 0 | 1.4 | 0 |
| Cit-Fibα$_{563-583}$ | 0 | 0 | 0 | 0 | 0 | 0 | 323.1 | 0 |
| Arg-Fibα$_{563-583}$ | 0 | 0 | 0 | 0 | 0 | 0 | 6.7 | 0 |
| Cit-Fibα$_{580-600}$ | 0 | 0 | 0 | 0 | 1.7 | 0 | 1.8 | 0 |
| Arg-Fibα$_{580-600}$ | 0 | 0 | 0 | 0 | 0 | 0 | 0.0 | 0 |
| Cit-Fibß$_{(Cit-R)62-81a}$ | 0 | 0 | 0 | 0 | 0 | 0 | 5.3 | 0 |
| Cit-Fibß$_{(R-Cit)62-81b}$ | 0 | 0 | 0 | 0 | 0 | 0 | 4.5 | 0 |
| Arg-Fibß$_{62-81}$ | 0 | 0 | 0 | 0 | 0 | 0 | 5.5 | 0 |
| CEP-1$_{5-21}$ | 0 | 0 | 0 | 0 | 0 | 0 | 47.3 | 0 |
| REP-1$_{5-21}$ | 0 | 0 | 0 | 0 | 0 | 0 | 0.0 | 0 |
| Ro52 | 6.7 | 6.9 | 9.2 | 5.96 | 0 | 0 | 1.9 | 0 |
| Jo-1 | 9.6 | 8.3 | | | 0.8 | 0 | 2.2 | 0.9 |
| PMScl100 | 2.2 | 6.98 | | | 1.7 | 0 | 2.7 | 0 |
| Scl70 | 18.4 | 0 | | | 1.0 | 0 | 2.4 | 0 |
| Sm | 0.0 | 0 | | | 0 | 0 | 2.2 | 0 |
| La(SS-B) | 12.4 | 10.2 | | | 1.1 | 0 | 1.6 | 0 |
| Ro60 | 4.6 | 5.9 | | | 0 | 0 | 1.3 | 0 |
| dsDNA | 0 | 0 | | | 0 | 0 | 3.5 | 0 |

*Weak   Strong*

*Not available*

*\* Some background staining on array*

**B**

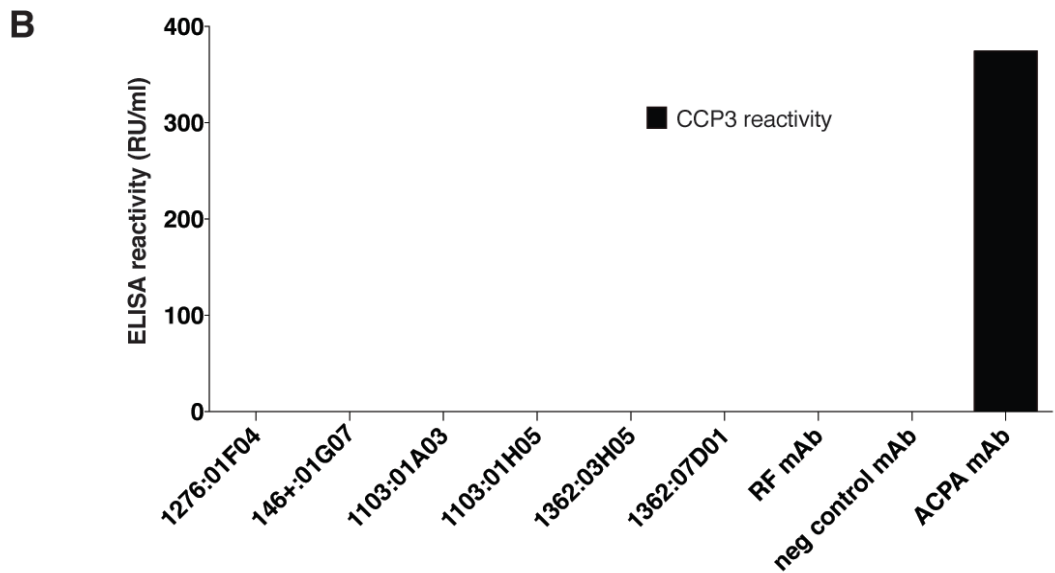

**Supplemental Figure 9. Synovial anti-MDA mAbs do not bind to citrullinated antigens**

**A.** Results from screened using Phadia's ImmunoCAP ISAC microarray system, where antibody binding to a large number of citrullinated peptides and their native arginine counterpart were compared (7). A selection of other autoimmunity-related antigens was also included. Activity Units (AU) are displayed in the heatmap, calculated form the fluorescence intensity by normalization to an internal control. **B.** ELISA screening of monoclonal IgG at 5ug/ml binding to CCP using the QUANTA Lite CCP3 assay (Inova Diagnostics). Values were compared to the ACPA mAb 1325:01B09 and the negative control 1325:01E02. Peptide information can be obtained in Supplemental Table 3.

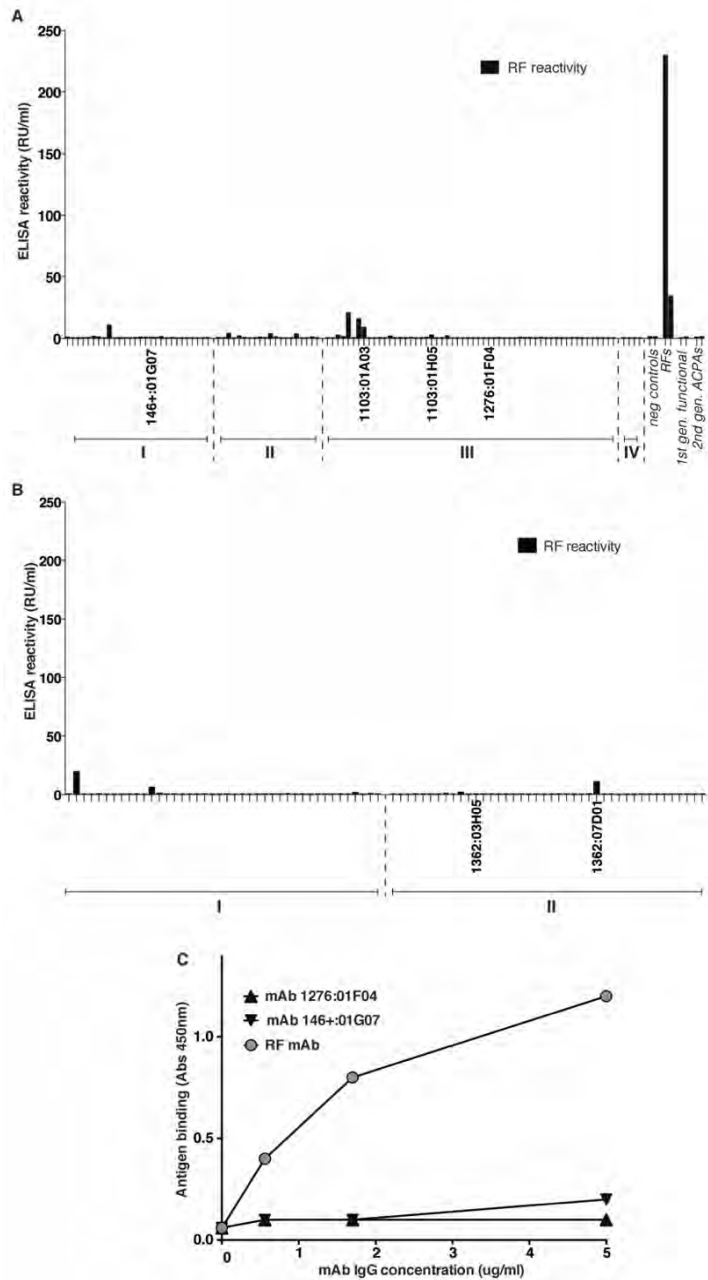

**Supplemental Figure 10. Synovial anti-MDA mAbs do not demonstrate any rheumatoid factor activity.**

Rheumatoid factor activity of the monoclonal antibodies was evaluated by ELISA using wells coated with rabbit IgG (3 ug/ml) and detection with rabbit (Fab')2 anti-human IgG (gamma-specific). Purified recombinant monoclonal IgG1 originated from **A.** synovial memory B cell clones, **B.** ASC clones, was tested at 5 ug/ml. Values for control mAbs (n=8) are shown in A. These included two negative controls (1276:01G09, 1362:01E02), two mAbs with RF activity (T13.146-B05, 1276:01C11), two first generation pathogenic functional mAbs (1276:01D10, 1103:01B02) and two second generation plasma cell derived ACPA mAbs (1325:01B09, 1325:03C03). **C.** RF reactivity of the anti-MDA memory B cell mAbs 1276:01F04, 146+:01G07 in serial dilution compared to the RF clone 1276:01C11 (also identified from synovial memory B cells).

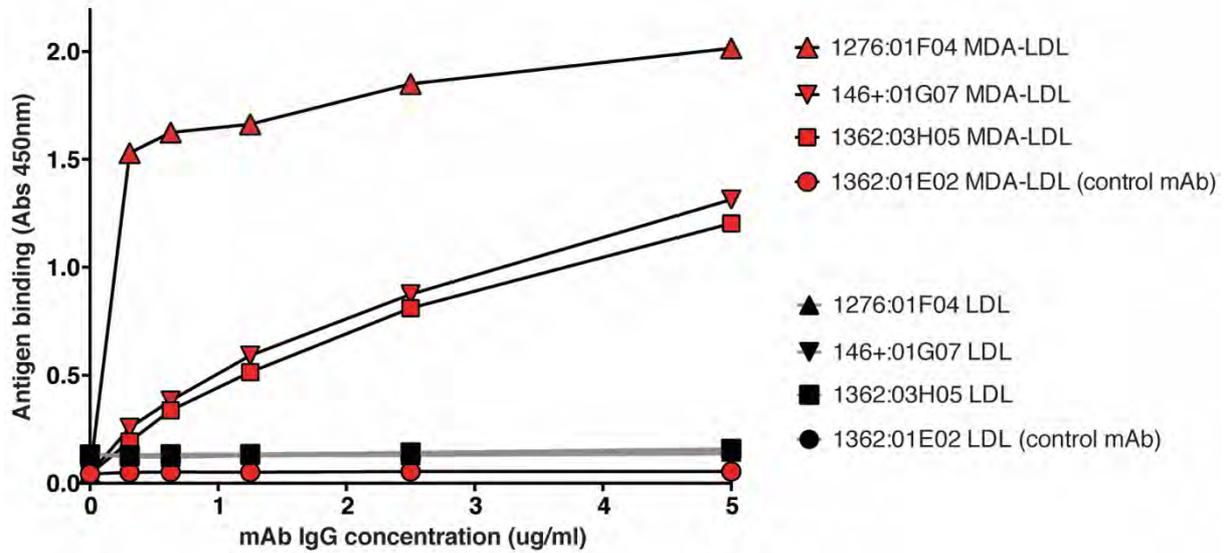

**Supplemental Figure 11. Synovial anti-MDA modified protein bind to MDA-LDL and not native LDL**

The figure depicts ELISA reactivity to MDA-modified human low density lipoprotein (MDA-LDL) compared to native unmodified LDL for three human synovial anti-MDA reactive IgG clones, 1276:01F04, 146+:01G07, 1362:03H05 and a negative control mAb 1362:01E02, at indicated concentrations.

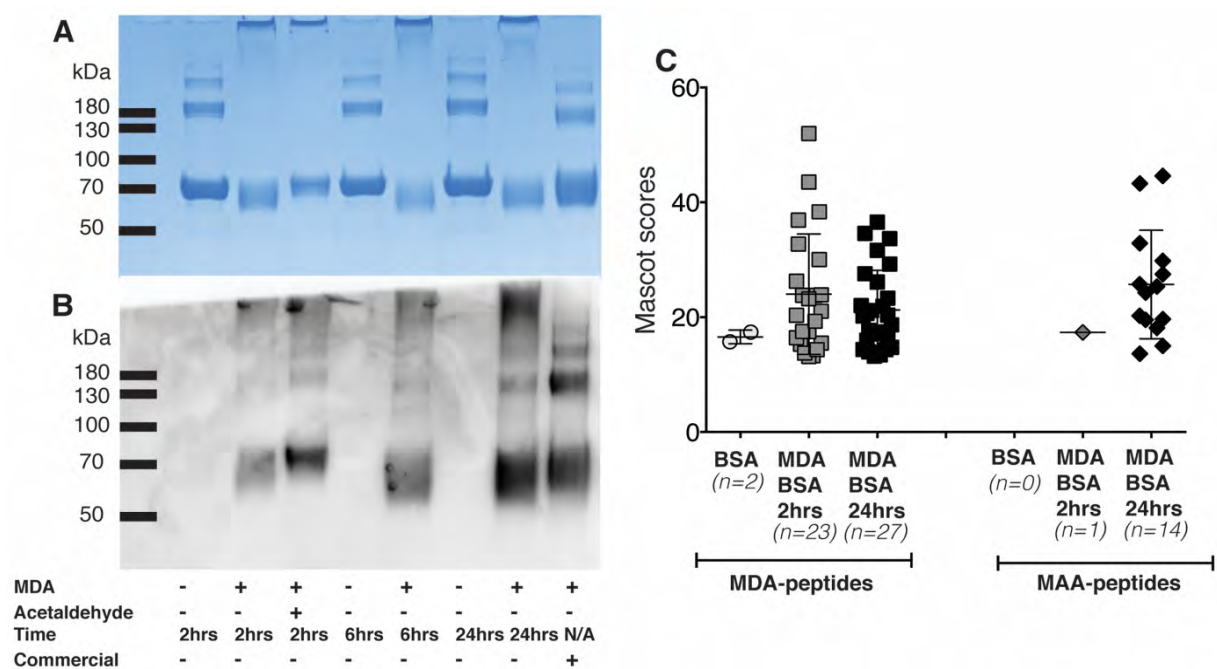

MDA   - + + - + - + +
Acetaldehyde   - - + - - - - -
Time   2hrs 2hrs 2hrs 6hrs 6hrs 24hrs 24hrs N/A
Commercial   - - - - - - - +

**Supplemental Figure 12. Different forms of MDA-modified BSA recognized by the anti-MDA mAb 1276:01F04**

In-house prepared malondialdehyde modified bovine serum albumin, MDA-BSA, was prepared by incubating molecular grade BSA with 50 mM MDA for 2hrs, 6hrs or 24hrs at 37C before extensive dialysis. Addition of 50% acetaldehyde in addition to MDA was used for generation of MAA-type modifications. Commercial MDA-MSA (far right lane) was used for comparison (Academy Bio-Medical). **A.** Coomassie blue stained SDS-PAGE (4-12% bis-tris gel with MES running buffer) of different lots of MDA-modified BSA. **B.** Western blot analysis of different lots of MDA-BSA stained with the monoclonal human recombinant antibody 1276:01F04 at 1 ug/ml. Control BSA incubated and treated the same way as modified BSA but without addition of MDA/acetaldehyde. The different BSA lots were denatured and reduced before run at 3 ug/lane. **C**. Results from mass spectrometry analysis of the untreated BSA compared to the BSA that had been MDA-treated for 2hrs or 24hrs. The graph shows the mascot scores for peptides identified to contain MDA-modified residues, MDA-lysine peptides (with N-ε-(2-propenal)lysine), MDA-modified arginine, or peptides that contained MAA-lysine peptides (with 4-methyl-1,4-dihydropyridine-3,5 dicarbaldehyde, MDHDC). Note that no acetaldehyde had been added to the reaction. The number of identified modified BSA peptides for each condition has been indicated. Only peptides with significant mascot score (>13) were included. Peptide sequences are presented in Supplemental Table 2.

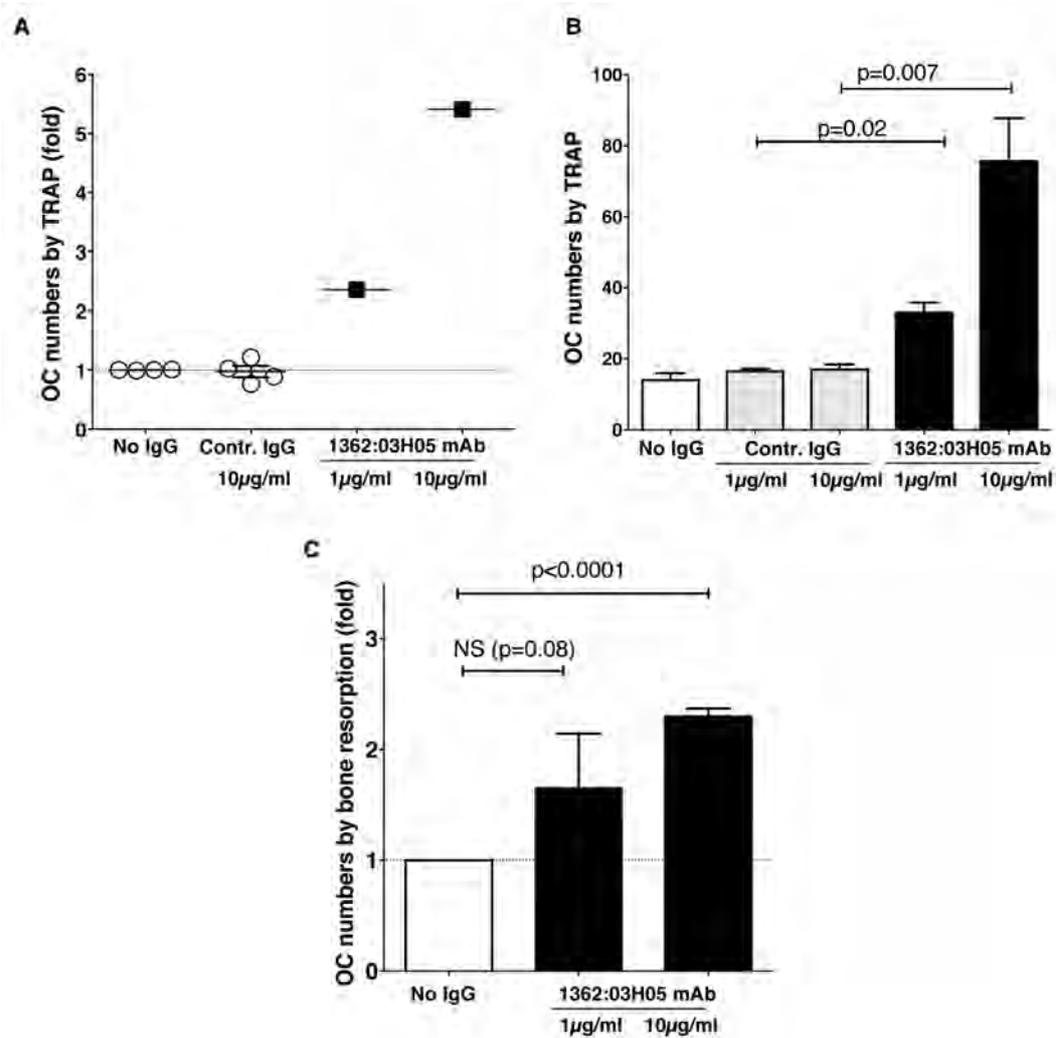

**Supplemental Figure 13. Plasma cell derived anti-MDA mAb enhance osteoclastogenesis**

The synovial plasma cell derived clone 1362:03H05 stimulates increased osteoclastogenesis in cultures of macrophage derived human osteoclasts. Osteoclasts were generated in vitro from CD14-positive monocytes isolated from the circulation of healthy donors and stimulated to induce osteoclast differentiation with M-CSF and RANKL with or without the presence of human purified monoclonal IgG1 at indicated concentrations. **A.** Fold increase osteoclasts measured by counts of tartate-resistant acid phosphatase (TRAP) positive cells with ≥3 nucleai. **B.** Numbers of TRAP positive cells. **C.** Fold increase of osteoclasts by resorption area on calcium phosphate plates. P-values are shown from 2-sided student's t-test. The mAb 1362:01E02 was used as negative control. The graph shows means of triplicates of a representative experiment.

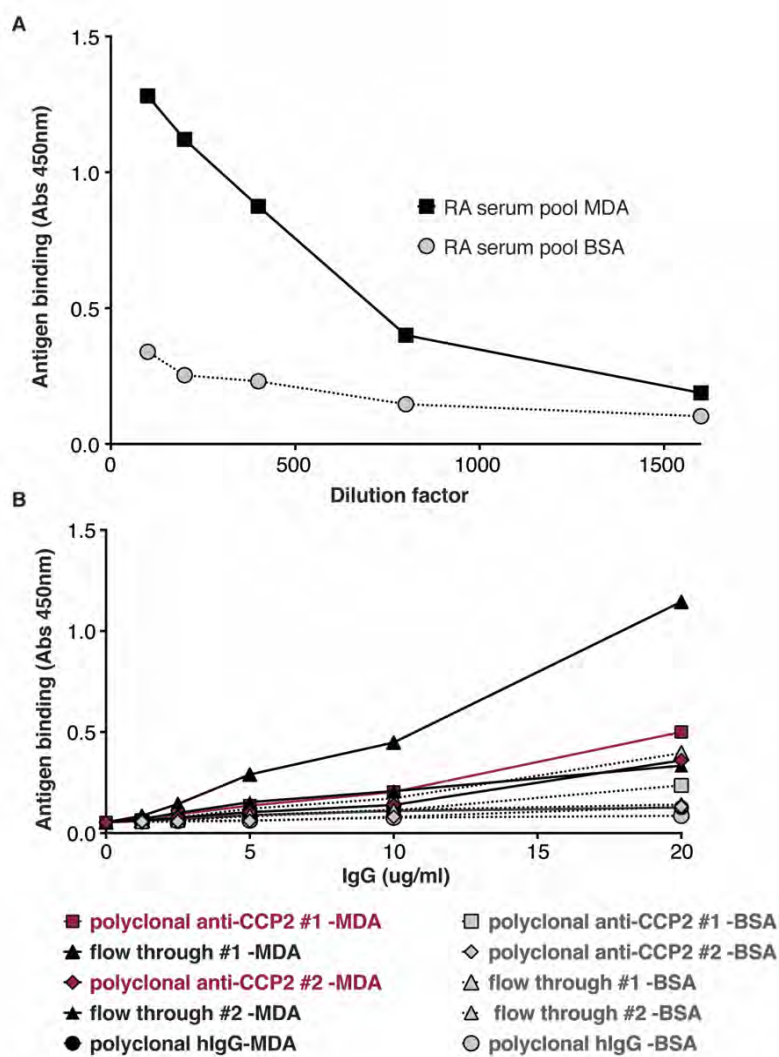

**Supplemental Figure 14. Polyclonal anti-CCP2 does not bind to MDA-modified protein**

The figure depicts ELISA data evaluating binding of polyclonal RA IgG to MDA-modified protein. **A.** Reactivity of pooled RA serum (14 seropositive RA patients), starting at dilution 1:100. **B.** Reactivity of affinity purified serum CCP2-reactive IgG from seropositive RA patients compared to the IgG flow through from the CCP2 column. Two independent batches are shown (#1 purified from 35 patients; #2 purified from 108 patients). Low or no binding above background was seen in the polyclonal anti-CCP2, some binding could be seen in one of the CCP2 negative fractions (flow through #1). Binding to MDA-BSA was compared to a control surface coated with only BSA. Chromopure human polyclonal IgG (Jackson ImmunoResearch) was used as negative control. Purification of anti-CCP2 IgG have been previously reported (8).